\documentclass[useAMS,usenatbib]{mn2e}
\usepackage{graphicx,amssymb,color}
\usepackage[normalem]{ulem}
\usepackage{url}

\newcommand{\rev}{ }
\newcommand{\revv}{ }

\title[Kuiper belts from formation to death]
{Linking the formation and fate of exo-Kuiper belts within solar system analogues}

\author[]{Dimitri Veras$^{1,2}$\thanks{E-mail: d.veras@warwick.ac.uk}\thanks{STFC Ernest Rutherford Fellow},
Katja Reichert$^3$, Francesco Flammini Dotti$^{4,5}$, Maxwell X. Cai$^{6,7}$, 
\newauthor
Alexander J. Mustill$^8$, Andrew Shannon$^9$,
Catriona H. McDonald$^{1,2}$, 
\newauthor
Simon Portegies Zwart$^7$,  M.B.N. Kouwenhoven$^4$,
Rainer Spurzem$^{3,10,11}$\thanks{Research Fellow of Frankfurt Institute for Advanced Studies}
\\
$^{1}$Centre for Exoplanets and Habitability, University of Warwick, Coventry CV4 7AL, UK
\\
$^{2}$Department of Physics, University of Warwick, Coventry CV4 7AL, UK
\\
$^{3}$Astronomisches Rechen-Institut, Zentrum f\"{u}r Astronomie der Universit\"{a}t Heidelberg, M\"{o}nchhofstr. 12-14, 69120 Heidelberg, Germany
\\
$^{4}$Department of Physics, School of Science, Xi'an Jiaotong-Liverpool University, 111 Ren'ai Rd., Suzhou Dushu Lake Science and 
\\
Education Innovation District, Suzhou Industrial Park, Suzhou 215123, P.R. China
\\
$^{5}$Department of Mathematical Sciences, University of Liverpool, Liverpool L69 3BX, UK
\\
$^{6}$SURFsara, Science Park 140, 1098 XG, Amsterdam
\\
$^{7}$Leiden Observatory, Leiden University, PO Box 9513, RA Leiden NL-2300, the Netherlands
\\
$^{8}$Lund Observatory, Department of Astronomy \& Theoretical Physics, Lund University, Box 43, 221 00, Lund, Sweden
\\
$^{9}$LESIA, Observatoire de Paris, Universit\'{e} PSL, CNRS, Sorbonne Universit\'{e}, Universit\'{e} de Paris, 5 place
\\
Jules Janssen, 92195, Meudon, France
\\
$^{10}$National Astronomical Observatories and Key Laboratory of Computational Astrophysics, Chinese Academy of Sciences,
\\
20A Datun Rd., Chaoyang District, Beijing 100101, China
\\
$^{11}$Kavli Institute for Astronomy and Astrophysics, Peking University, Yiheyuan Lu 5, Haidian Qu, Beijing 100871, China 
}

\pubyear{2019}

\begin{document}
\label{firstpage}
\pagerange{\pageref{firstpage}--\pageref{lastpage}}
\maketitle

\begin{abstract}
Escalating observations of exo-minor planets and their destroyed remnants both passing through the solar system and within white dwarf planetary systems motivate an understanding of the orbital history and fate of exo-Kuiper belts and planetesimal discs. Here we explore how the structure of a $40-1000$~au annulus of planetesimals orbiting inside of a solar system analogue that is itself initially embedded within a stellar cluster environment varies as the star evolves through all of its stellar phases. We attempt this computationally challenging link in four parts: (1) by performing stellar cluster simulations lasting 100~Myr, (2) by making assumptions about the subsequent quiescent 11 Gyr main-sequence evolution, (3) by performing simulations throughout the giant branch phases of evolution, and (4) by making assumptions about the belt's evolution during the white dwarf phase. Throughout these stages, we estimate the planetesimals' gravitational responses to analogues of the four solar system giant planets, as well as to collisional grinding, Galactic tides, stellar flybys, and stellar radiation. We find that the imprint of stellar cluster dynamics on the architecture of $\gtrsim100$ km-sized exo-Kuiper belt planetesimals is retained throughout all phases of stellar evolution unless violent gravitational instabilities are triggered either (1) amongst the giant planets, or (2) due to a close ($\ll 10^3$ au) stellar flyby. In the absence of these instabilities, these minor planets simply double their semimajor axis while retaining their primordial post-cluster eccentricity and inclination distributions, with implications for the free-floating planetesimal population and metal-polluted white dwarfs. 
\end{abstract}


\section{Introduction}

The Kuiper Belt and scattered disc refer to the collection of minor planets which orbit the Sun at separations between tens and hundreds of astronomical units. These objects provide crucial constraints on the temporal evolution of the solar system. With respect to formation, the {\it New Horizons} mission 
has revolutionised our understanding of Pluto's geophysical history \citep{steetal2015,mooetal2016} as well as the formation pathway of 
the ``squashed snowman'' binary object 486958 Arrokoth (2014 MU69 Ultima Thule) \citep{nesetal2019,steetal2019}. Alternatively, with respect to the fate of Kuiper Belt
and scattered disc objects, we must look towards the interstellar interlopers or \emph{s\={o}lus lapis} 1I/‘Oumuamua \citep{meeetal2017,williams2017} and 2I/Borisov
\citep{guzetal2019} or elsewhere, in evolved extrasolar planetary systems \citep{2018MNRAS.479L..17P}.

Besides the Sun, the only other stars around which the presence of individual minor planets have been observed or inferred are young hot stars -- where variable absorption is inferred to come from gas \citep{ferletetal1987,welshmontgomery2015} and possibly dust \citep{rappaportetal2018} released by individual comets -- 
 and white dwarfs \citep{vanetal2015,manetal2019,vanetal2019}, which represent the endpoint of stellar evolution for nearly all Milky Way stars.
The destroyed remnants of other minor planets are observed through the presence of over 40 debris discs around white dwarfs
\citep[e.g.][]{zucbec1987,graetal1990,gaeetal2006,farihi2016,denetal2018,swaetal2019a} and the constituents of
over 1000 minor planets are observed inside the atmospheres of white dwarfs \citep{zucetal2003,zucetal2010,koeetal2014,couetal2019}. 
This minor planet destruction provides a unique means to probe the bulk 
composition of asteroids and comets beyond our Solar System.

Consequently, linking the formation and fate of minor planets has importance in multiple astrophysical contexts. 
However, the computationally challenging nature of this task has motivated investigations which focus on a single or a few
phases of stellar evolution. For example, the dynamical origin of the currently observed Kuiper Belt architecture has been the subject of numerous
investigations \citep[e.g.,][]{gometal2008,liokau2008,2014MNRAS.444.2808P,nesvorny2018} that are almost entirely restricted to the early main-sequence
phase of the Sun, or have considered just the post-formation evolution \citep[e.g.,][]{tiscarenoetal2009,lawleretal2017,shadaw2018}. Other theoretical studies dedicated to minor planets are focused solely on the white dwarf phase 
\citep[e.g.,][]{alcetal1986,wyaetal2014,stoetal2015,veretal2015a,veretal2016a,broetal2017,griver2019,makver2019}, 
and sometimes on individual systems {\rev \citep{guretal2017,veretal2017a,duvetal2019,veretal2020a}}.

Between the main-sequence and white dwarf phases of stellar evolution, most stars experience violent physical alterations
during the giant branch phases. Minor planets are particularly susceptible to these changes \citep{veras2016a} and the consequences
have again represented the focus of investigations during this phase alone. These changes include radiative destruction due to YORP-induced rotational
fission \citep{veretal2014a,versch2019} -- potentially contributing to the formation of debris discs \citep{bonwya2010,bonetal2013,bonetal2014} --  and orbital evolution due to the both the Yarkovsky effect \citep{veretal2015b,veretal2019}
and stellar mass loss \citep{omarov1962,hadjidemetriou1963}.  In particular, stellar mass loss can eject these minor planets \citep{veretal2011}, which may eventually pass through other planetary systems \citep{doetal2018,moromartin2019}.

Other investigations which included minor planet evolution have attempted to bridge the gap between several stellar 
evolutionary phases and/or included the influence of major planets  
\citep{bonetal2011,debetal2012,frehan2014,veretal2016b,caihey2017,musetal2018,smaetal2018,smaetal2019,antver2016,antver2019}  
or binary stars \citep{bonver2015,hampor2016,petmun2017}.  None, however, have calibrated their minor planet initial conditions
-- which could sensitively determine the eventual accretion rate onto white dwarfs -- 
with the results of planetary system formation within stellar clusters.

Nearly all planetary systems are formed in clustered stellar environments. Hence, theoretical investigations
of these systems are of paramount importance, despite detections
of only a few dozen cluster planets (\citealt*{bruetal2017}, \citealt*{manetal2017}, \citealt*{leaetal2018}, and see Table~1 of \citealt*{caietal2019} for an up-to-date listing of cluster planets). Many studies
have focused on how photoevaporation from young hot stars affects major planet formation and migration 
\citep{adaetal2004,verarm2004,andetal2013,daietal2018,winetal2018,conetal2019,nicetal2019} and
how frequent and slow stellar flybys influence the resulting major planet architecture 
\citep{maletal2007,maletal2011,haoetal2013,zheetal2015,shaetal2016,hamtre2017,caietal2017,caietal2019,flaetal2019,lietal2019,flybyetal2019}. Less-studied
have been the consequences for minor planets {\rev \citep[see, e.g.,][]{brasseretal2006,pfaetal2018,hanetal2019,batetal2020}}. 
   
\begin{figure*}
\includegraphics[width=\textwidth]{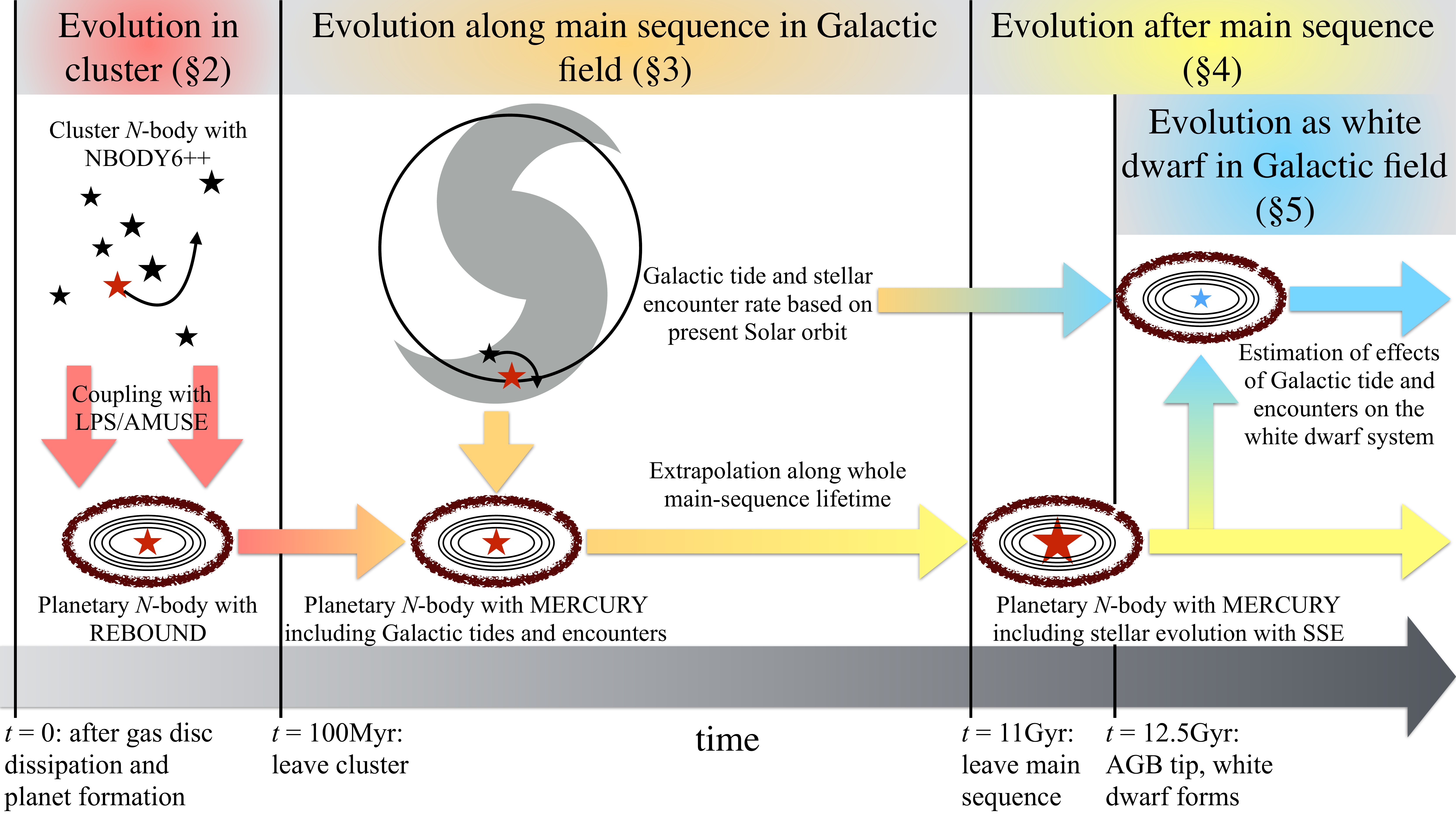}
\caption{Overview of the evolution of our planetary 
systems and the numerical codes (in all caps) used to study each stage. 
In Section~\ref{sec:cluster} we place a newly-formed 
planetary system in a stellar cluster and follow its evolution for 100\,Myr. We then take the resulting system configuration at 100\,Myr -- assumed to be the time at which the cluster dissolves -- and integrate the planetary 
and planetesimalal orbits under the effects of stellar and planetary gravity, Galactic tides and 
encounters with stars in the field (Section~\ref{sec:MS}). Because we find that the effects 
of tides and flybys are unimportant, and the system is inherently stable, we assume that the system configuration at the end of the main sequence -- at 11 Gyr -- remains unchanged from the configuration at 100 Myr. We then use this 
configuration as initial conditions for our post-main-sequence simulations, which feature red giant branch and asymptotic giant branch (AGB) mass loss, followed by several Gyr of white dwarf cooling (Section~\ref{sec:postMS}). Finally, we also estimate the effects of the Galactic environment on the expanded 
systems orbiting white dwarfs (Section~\ref{sec:WD}).}
\label{fig:cartoon}
\end{figure*}

Here, we attempt to qualitatively trace the evolution, across all stellar phases, of exo-Kuiper Belts and planetesimal discs which are shaped 
in a stellar cluster environment. In order to restrict the enormous parameter space associated with this task, we consider only
solar system analogues (Sun-like stars with Jupiter, Saturn, Uranus and Neptune) and split the temporal evolution into four distinct
segments: [i] stellar cluster evolution (Section 2), [ii] main-sequence evolution (Section 3), [iii] giant branch evolution (Section 4) and
[iv] white dwarf evolution (Section 5). 
A schematic overview of the different segments and 
the numerical codes used is given in Figure~\ref{fig:cartoon}.
We discuss the results in Section 6 and conclude in Section 7.

\section{Star cluster evolution}
\label{sec:cluster}

Star formation occurs in regions with a stellar density that is substantially higher than that of the Galactic neighbourhood. 
Consequently, within a cluster, close stellar flybys are frequent and slow; they can help sculpt the orbital distribution of objects which are sufficiently far from their parent stars. Modelling these interactions is challenging primarily because of the different timescales (cluster evolution and planetary evolution) which must be simulated self-consistently, and secondarily because of the large parameter space.

In this section, we detail our cluster simulations. We describe our computational approach in Section 2.1 before outlining the initial conditions for our star clusters (Section 2.2) and planetary systems (Section 2.3); we report the simulation output in Section 2.4.

\subsection{Computational approach}

Planetary system evolution is fundamentally different from star cluster evolution. In the absence of major instabilities, planets evolve quiescently and are subject to only minor perturbations from mutual gravitational interactions and external perturbations from passing stars. The timescale for orbital changes is typically secular.

Star clusters, on the other hand, evolve through two-body relaxation and close few-body encounters (which may involve close binaries). These processes produce changes on orbital timescales rather than secular timescales. Further, star cluster dynamics more readily exhibits deterministic chaos, where systems with slightly different initial conditions diverge from each other in phase space exponentially, in less than an orbital time \citep[e.g.,][]{mil1964,quitre1992}. 

For these reasons, executing a combined simulation of planetary systems in star clusters is challenging.
The main obstacle to overcome is not actually establishing the different timescales or hierarchies in the integrations\footnote{Close stellar binaries and planetary systems can be treated similarly.}, but rather accurately modelling the resonant and secular effects in the internal evolution of planetary systems. While different authors have advocated special symplectic methods to model planetary systems accurately \citep[e.g.][]{wisdom1991},  \cite{koketal1998} demonstrated how small improvements in the Hermite integrator used for star cluster simulations can lead to an accurate treatment of its internal planetary systems over Gyr timescales.

We previously successfully performed simulations involving star clusters and their internal internal planetary systems by using a single $N$-body code \citep[\texttt{NBODY6++GPU} with massless particles, e.g., Shu et al. in preparation,][]{spuetal2009}. In this work, we follow a different method, one used by \cite{flaetal2019} and \cite{caietal2017,caietal2019}. Here we first integrate the star cluster dynamics, recording the trajectories of stars and their nearest neighbours. Then, we add in planetary systems (including the exo-Kuiper belt objects) to selected host stars; the evolution of these planetary systems is performed by another code, named \texttt{LonelyPlanets} {\revv \citep{caietal2017, caietal2018, caietal2019, flaetal2019,flaetal2020}}. \texttt{LonelyPlanets} utilises the recorded perturbations from the star cluster simulation outputs as external perturbations to the planetary systems.

Hence, the star cluster simulations are carried out by \texttt{NBODY6++GPU} \citep{wang2015b, wang2016}, whereas the planetary systems are integrated by \texttt{LonelyPlanets} (hereafter, \texttt{LPS}). This approach represents a fast and accurate method for integrating the evolution of planetary systems in star clusters, under the condition that the force of the planets on the stellar population can be neglected.
The star cluster evolution is integrated through the Hermite scheme, using \texttt{NBODY6++GPU}. This code is the latest updated version of the original \texttt{NBODY6} \citep{aarseth1999} and \texttt{NBODY6++} \citep{spurzem1999}. The greatest improvement in the last version is the feature which takes advantage of  graphical processing units (GPUs) and task parallelization. The latter is achieved through MPI \citep[Message Passing Interface;][]{tapamo2009}, where both regular and irregular forces are parallelized. The GPU usage significantly improves \texttt{NBODY6++GPU} performance, especially for  long-range (regular) gravitational forces

\texttt{LPS} is based on the \texttt{AMUSE} framework \citep{portegieszwart2011,mcmillan2012,pelupessy2013, portegieszwart2018}. \texttt{AMUSE} helps us consolidate our methodology, which can be summarised in four steps: (i) setting initial conditions for modelling the star cluster and planetary systems; (ii) modelling numerically the dynamics and stellar evolution of the star cluster; (iii) identifying close encounters experienced by the planet-hosting stars; {\rev and (iv) modelling the evolution of the planetary systems under the influence of the closest perturbers. 
In our work we will consider the five closest perturbers} to the host stars, and these perturbers are identified by using the standard method for neighbour selection in the \texttt{NBODY6++GPU} simulations.

\texttt{NBODY6++GPU} integrates the star clusters, and stores data at a high temporal resolution (i.e., star cluster output) by using the block time step scheme \citep[BTS,][]{caietal2015}. The latter approach prevents data redundancy for planetary systems in the low temporal resolution regime (i.e., planetary output). These data are then sent to the \texttt{REBOUND} integrator \citep{rei2012}, which integrates the planetary system until the next BTS output set of data: in our dataset the star cluster output time-step of $\sim 1000$ years was adopted. {\rev We use the HDF5\footnote{https://www.hdfgroup.org/} output format \citep{portell2011}, a highly efficient storage scheme organised in a database-like structure.}

As the star cluster and the planetary systems evolve, escaping bodies are removed. The criterion for escape of a star in a stellar cluster in \texttt{NBODY6++GPU} is when $r > 2r_\mathrm{tid}$, where $r_\mathrm{tid}$ is the star cluster's tidal radius (see Sec. \ref{sec:star_cluster} for further info). In \texttt{LPS}, the escapers from the planetary system are those particles with orbital eccentricities $e > 0.995$.

\subsection{Star cluster initial conditions}\label{sec:star_cluster}

We place our planetary systems each within a stellar cluster of 2000 stars containing an initial total mass of $1139$~M$_{\odot}$. The initial mass function (IMF) for the stars in the cluster follows a \cite{kroupa2001} IMF. The stellar masses are drawn from the range $0.08-100$~M$_\odot$ with an expected average mass of $0.57$~M$_\odot$. We use a \citep{plummer1911} model in virial equilibrium as the density profile for the cluster from which the initial positions and velocities of the stars are drawn. The initial half-mass radius $r_\mathrm{hm}$ is $0.73$~pc while the initial central density is $896$~M$_\odot/\mathrm{pc}^3$. These values are motivated from studies of the star cluster in which the Sun might have been born \citep{2009ApJ...696L..13P,2019A&A...622A..69P}. Furthermore, we assume a standard solar neighborhood tidal field \citep{heitre1986} %\citep{caietal2016} 
and no primordial mass segregation. 

{\revv The fraction of primordial binary systems is set to $0$ per cent in order to (i) keep the external perturbations ``clean" and associate external perturbations entirely with single stars and not three-body effects, (ii) avoid circumbinary planetary systems and potential exchange interactions, and (iii) maintain a reasonable computational cost for the simulations. In reality, observations indicate that the binary fraction varies from a few per cent to half, where the lower bound is more representative of globular clusters and the upper bound is more representative of the field \citep{geletal2008,ragetal2010,miletal2012}. A binary flyby typically increases the cross-section for planetary ejection and collisions \citep{wanetal2020}. Hence, including binary flybys would have likely increased the extent of destabilisation of the disc, although the severity of the destabilisation would have been dependent on multiple parameter choices.}
%the destabilization in the disc, although the extent to which would have been dependent on multiple parameter choices.}

\texttt{NBODY6++GPU} includes stellar evolution of single and binary stars according to \cite{huretal2000} and \cite{huretal2001}. We assume a solar metallicity for all stars, and include further improvements by taking into account fallback from supernova explosions and stellar winds of massive stars \citep{beletal2002}. Also, kicks occurring at the formation of neutron stars according to \cite{hobetal2005} are included. Although stellar evolution is modelled in our cluster simulations, the effects of mass loss of the host star on the planetary systems are not: the consequences for the major and minor planets during the main sequence phase would be negligible for our 1~M$_\odot$ host stars. 

We define a tidal radius $r_\mathrm{tid}$  as given in \cite{caietal2016}. This radius is similar to the Jacobi radius $r_\mathrm{J}$ and is related to the quantity $r_{\mathrm{lim}}$ that was defined by \cite{kin1962} (see \citealt*{ernetal2009} for further discussion). The differences between these three quantities are small, of order unity. Our $N$-body simulation considers stars who reach twice the tidal radius as escapers and removes them from the system, meaning that some stars which are unbound from the cluster are followed. We do acknowledge, however, that the concept of a spherical tidal radius is an approximation; in reality the tidal escape problem is much more complicated (cf. e.g. \citealt*{ernetal2009}). 
By keeping stars in the simulation which are not bound to the cluster, we can follow for the dynamics near the Lagrangian points with our $N$-body simulations in detail.

\begin{figure*}
\includegraphics[width=\textwidth]{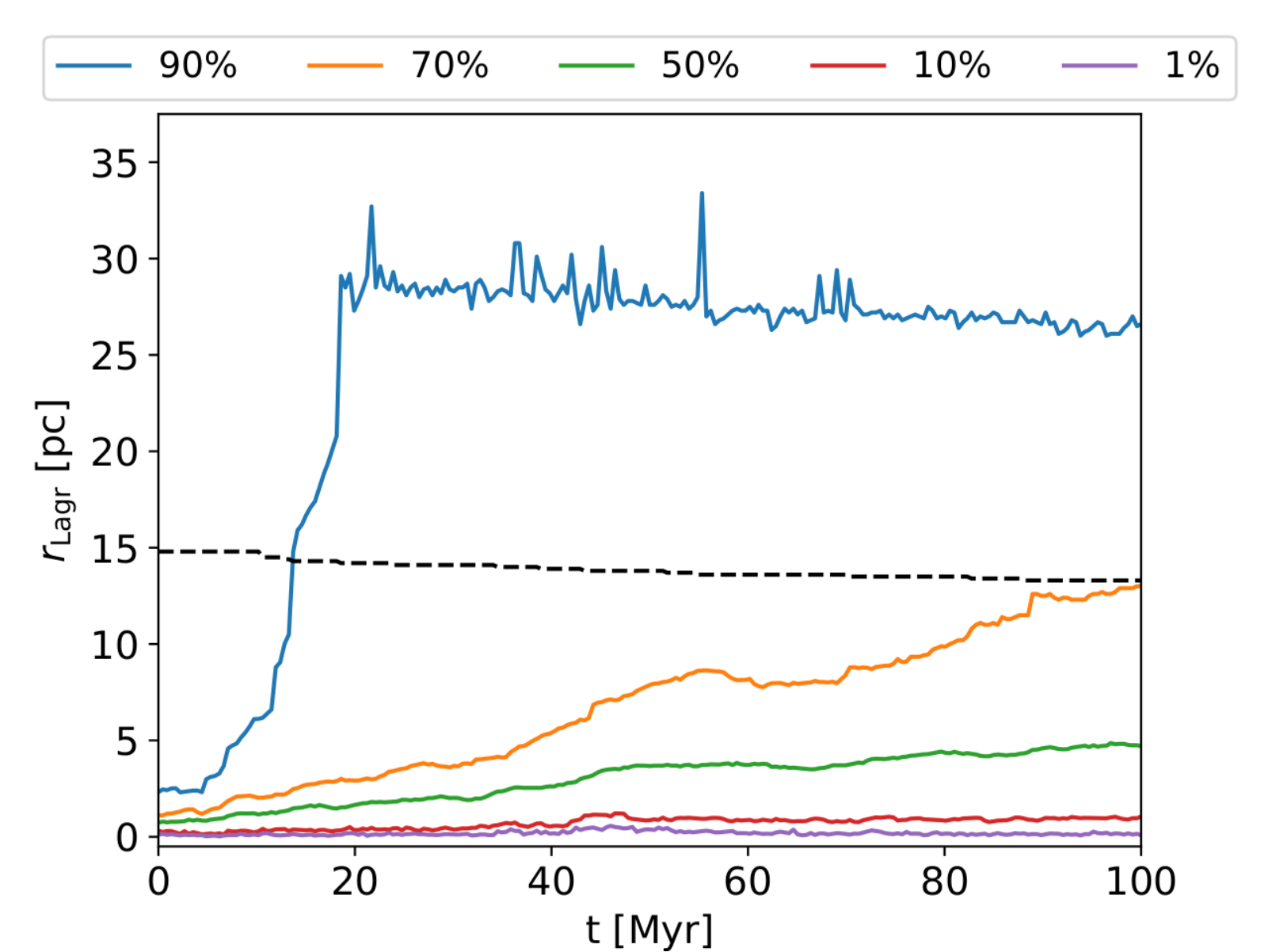}
\caption{
Lagrangian radii $r_\mathrm{Lagr}$ of the star cluster, containing the fraction of current total mass (from 1-90 per cent, from the bottom curve moving upwards;  see legend on top of plot), as a function of time. For comparison, the tidal radius $r_\mathrm{tid}$ is shown by the black dashed line. See main text for further explanation.
}
\label{fig:rlagr}
\end{figure*}

Fig.~\ref{fig:rlagr} provides an overview of the global evolution of our star cluster through the Lagrangian radii, which contain the indicated fraction of current total mass. Note that the 90\% Lagrangian radius stays approximately constant after it has reached twice the tidal radius, because at this point we start removing particles from the simulation. We have also plotted the tidal radius as a function of time in that figure. At the time where the tidal radius is equal to the half mass radius (50\% Lagrangian radius), the mass loss of the bound part of the star cluster is 50\%: this time defines the half-life of the cluster. In this paper we simulate the cluster environment of our planetary systems for 100 Myr because after this time the cluster is going to dissolve and will have reduced its central density significantly; most strong encounters will have occurred by 100 Myr.

%%%%%%%%%%%%%%%% Figure 
%%%%%%%%%%%%%%%% Figure 

\subsection{Planetary system initial conditions}

Now we turn to the planetary systems. We define $t=0$ as the time when the natal protoplanetary disc has dissipated, and all that remains are major and minor planets, as well as the parent Sun-like star (with a mass of 1.0~$\mathrm{M}_{\odot}$). 

At this time, we emplace four major planets {\rev into an} artificial system. The major planets are analogues of Jupiter, Saturn, Uranus and Neptune in mass, radius and current separation.  
As the protoplanetary disk lifetime was likely a few million years \citep{pascuccitachibana2010,williamscieza2011}, and (at least in the Sun's case) the protostar may form a few Myrs after the first stars in the cluster \citep{adams2010}, the star may depart the cluster slightly earlier than is assumed here.

Besides circular, co-planar orbits, we assume currently-observed planetary separations because the original orbital distribution of the four giant planets in our solar system is unknown. The giant planets were likely in a more compact configuration in the past, and Neptune's outward migration explains resonant populations in the Kuiper Belt \citep{Malhotra1993}. On the other hand, a prior inwards migration of the giant planets can explain Jupiter's composition and the asymmetry of its Trojan populations \citep{ObergWordsworth2019,Pirani+19}. Additionally, the giant planets likely underwent at least one gravitational instability which has re-ordered the planets at some, still-uncertain, previous epoch \citep{thoetal2002,tsietal2005,moretal2007,nesmor2012,morbidelli2018}. These instabilities have also
shaped the Kuiper Belt and scattered disc. Furthermore, the current configuration of the four giant planets in the solar system is not expected to undergo any future instability amongst themselves (without an external influence) during the remainder of the Sun's main-sequence lifetime \citep{lasgas2009,hayetal2010,zeebe2015,veras2016b}.

This analogue solar system serves as a template for our simulations. We generate {\rev 11} of these systems and impose different planetesimal disc configurations on each. We deposit 2000 massless particles in each planetesimal disc in an annulus extending from $a=40$~au to 1000~au, placing them uniformly in semimajor axis. {\rev Although these test particles are massless, we treat each as having a diameter of at least 100 km for the later application of radiative forces. The outer distance bound of 1000~au approximates the aphelion of distant solar system objects such as 90377 Sedna.} These particles feel the gravitational force of the star and planets, but do not exert a force. The particles, which will henceforth be denoted by planetesimals, are on initially circular and co-planar orbits with randomly distributed mean anomalies. 

These discs are initially dynamically cold in order to ensure that subsequent changes to the eccentricity and inclination are not primordial, and can be interpreted entirely in terms of gravitational excitation by stars and major planets. These discs are similar to the broad planetesimal discs formed in the models of \cite{Carrera+17}, where photo-evaporation of a disc several hundred au wide drives efficient planetesimal formation by the streaming instability at large radii.

{\rev However, there is no observational confirmation that the models adopted in \cite{Carrera+17} are correct. Hence, because the disc is composed of test particles, we could divide the disc into different regions that may be treated independently. One natural dividing line is the maximum radial extent of observed debris discs. This value is debatable, particularly with respect to how they evolve out of protoplanetary discs \citep{andrews2020}, but we adopt a value of 150 au. 

A value of 150 au appears to represent a reasonable upper limit given infrared observations of debris discs \citep[][and the catalogue of resolved debris discs by Pawellek and Krivov\footnote{\url{https://www.astro.uni-jena.de/index.php/theory/catalog-of-resolved-debris-disks.html}}]{sibetal2018}; see also \cite{krietal2013} with the caveat that some of these discs may actually be background galaxies \citep{gasrie2014}. However, we caution that  non-detectability does not necessarily translate into absence, because dust production drops off inversely to the third or fourth power of distance \citep{shawu2011}. Nevertheless, we henceforth characterize our discs in two regions, from (i) $40-150$ au, and (ii) $40-1000$ au.
}

% It might be more physically motivated to assume a Rayleigh distribution (see Lissauer \& Stewart 1993; Kokubo \& Ida 2002), but for simpilcity I just assumed zero initial inclination and eccentricity.

\subsection{Results from star cluster simulations}

The key results of these simulations are: (i) flybys are rarely intrusive enough to non-negligibly alter the orbital parameters of the major planets, (ii) the resulting distributions of the planetesimal discs differ substantially between systems, reflective of the perturbative environment in which they were placed, {\rev and (iii) the observed debris disc range ($40-150$) au is well-protected from cluster perturbations except in extreme cases}. For example, the fraction of surviving planetesimals after 100 Myr in the 11 systems span almost the entire possible range; see Table \ref{tab:loss} {\rev for the $40-1000$~au sample, and Table \ref{tab:loss2} for the $40-150$~au sample}.

\begin{table}
	\centering
	\caption{The number and per cent of planetesimals {\rev from an initial annulus of $40-1000$ au} remaining after 10 Myr, 90 Myr and 100 Myr in our cluster simulations. The four highlighted systems in the first column are used in separate figures throughout the paper.}
	\label{tab:loss}
	\begin{tabular}{ccccc} % four columns, alignment for each
		\hline
		 Highlighted & & & &
		 \\
		systems & 0 Myr & 10 Myr & 90 Myr & 100 Myr\\
		\hline
		& 2000 & 2000 & 2000 & 2000 (100\%)\\ %Dataset 108
		& 2000 & 2000 & 1996 & 1996 (99.8\%)\\ %Dataset 105
		System \#3 & 2000 & 1989 & 1963 & 1963 (98.2\%)\\ %Dataset 122
		System \#4 & 2000 & 1985 & 1958 & 1957 (97.9\%)\\ %Dataset 110
		& 2000 & 2000 & 1645 & 1641 (82.1\%)\\  %Dataset 100
		& 2000 & 2000 & 1515 & 1507 (75.3\%)\\  %Dataset 120
		& 2000 & 1714 & 961 & 957 (47.9\%)\\  %Dataset 125
		System \#1 & 2000 & 1832 & 853 & 853  (42.7\%)\\  %Dataset 11
		& 2000 & 1985 & 725 & 719 (36.0\%)\\  %Dataset 157
		System \#2 & 2000 & 1221 & 193 & 193 (9.7\%)\\  %Dataset 15
		& 2000 & 917 & 60 & 57 (2.9\%)\\  %Dataset 109
		\hline
	\end{tabular}
\end{table}

\begin{table}
	\centering
	\caption{{\rev The number and per cent of planetesimals from an initial annulus of $40-150$ au remaining after 10 Myr, 90 Myr and 100 Myr in our cluster simulations. The system order listed is the same as in Table \ref{tab:loss}, even though here the per cent of escaped systems is not monotonically decreasing.}}
	\label{tab:loss2}
	\begin{tabular}{ccccc} % four columns, alignment for each
		\hline
		 Highlighted & & & &
		 \\
		systems & 0 Myr & 10 Myr & 90 Myr & 100 Myr\\
		\hline
		& 225 & 225 & 225 & 225 (100\%)\\ %Dataset 108
		& 225 & 225 & 225 & 225 (100\%)\\ %Dataset 105
		System \#3 & 225 & 225 & 225 & 225 (100\%)\\ %Dataset 122
		System \#4 & 225 & 225 & 225 & 225 (100\%)\\ %Dataset 110
		& 225 & 225 & 225 & 225 (100\%)\\  %Dataset 100
		& 225 & 225 & 225 & 225 (100\%)\\  %Dataset 120
		& 225 & 225 & 200 & 200 (88.9\%)\\  %Dataset 125
		System \#1 & 225 & 225 & 225 & 225  (100\%)\\  %Dataset 11
		& 225 & 225 & 225 & 225 (100\%)\\  %Dataset 157
		System \#2 & 225 & 225 & 153 & 153 (68.0\%)\\  %Dataset 15
		& 225 & 225 & 26 & 26 (11.6\%)\\  %Dataset 109
		\hline
	\end{tabular}
\end{table}

%%%%%%%%%%%%%%%% Figure (datasets 11 and 15)
\begin{figure*}
{\Large\bf Cluster Evolution: System \#1
\ \ \ \ \ \ \ \ \ \ \ \ \ \ \ \ \ \
\Large\bf Cluster Evolution: System \#2}
\centerline{
\includegraphics[width=8.5cm]{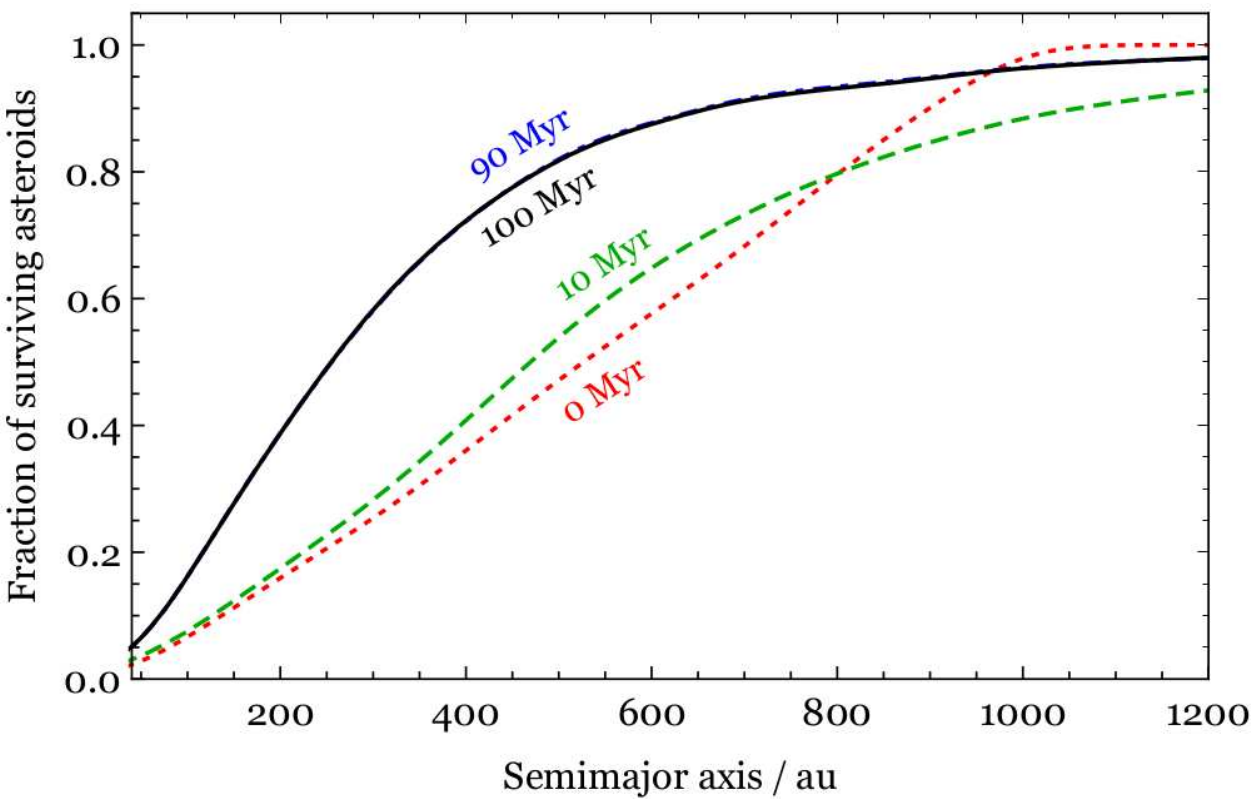}
\ \ \ \ \ \
\includegraphics[width=8.5cm]{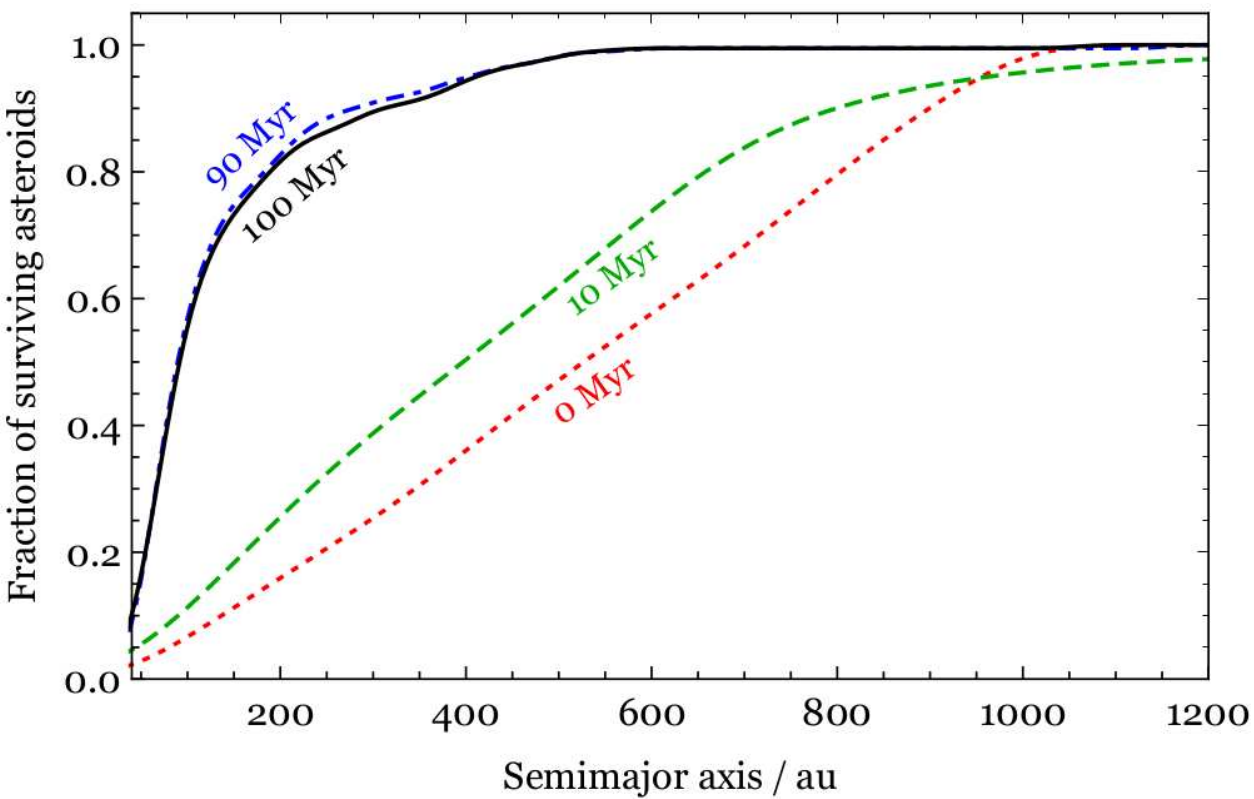}
}
\centerline{}
\centerline{
\includegraphics[width=8.5cm]{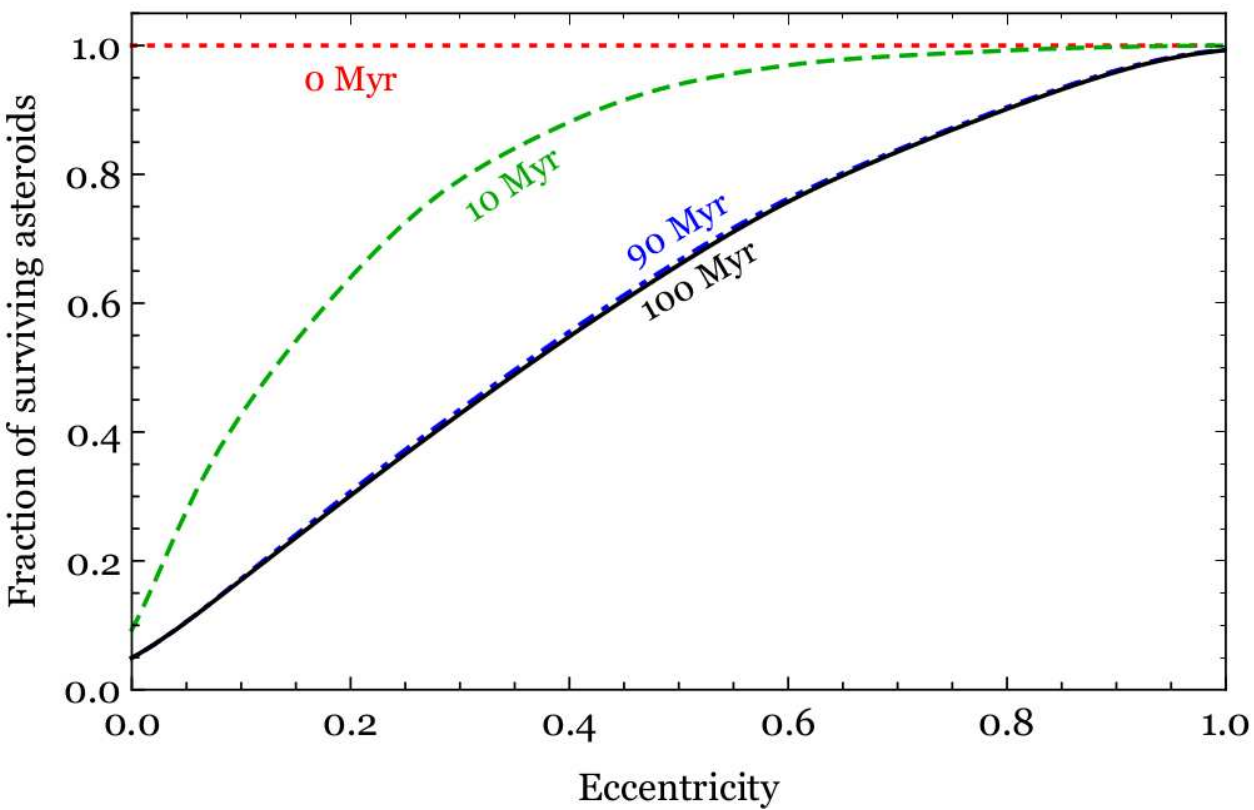}
\ \ \ \ \ \
\includegraphics[width=8.5cm]{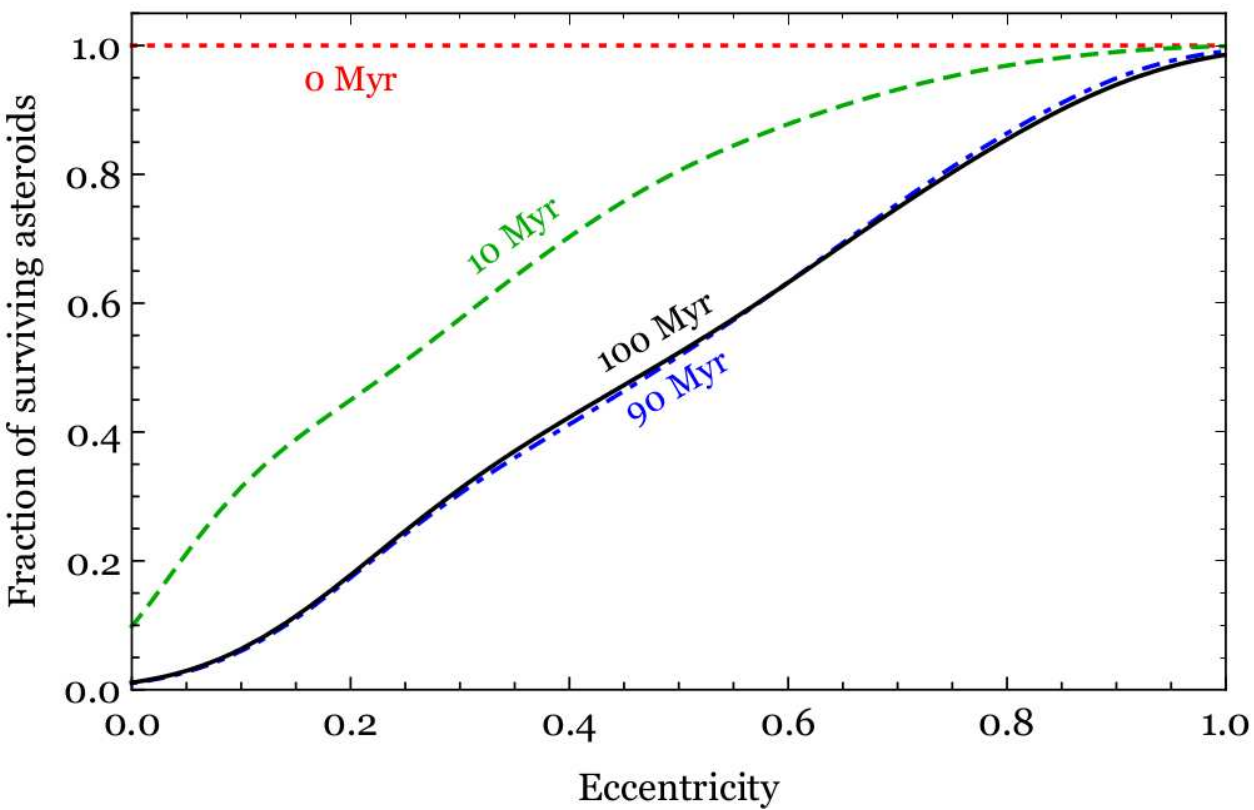}
}
\centerline{}
\centerline{
\includegraphics[width=8.5cm]{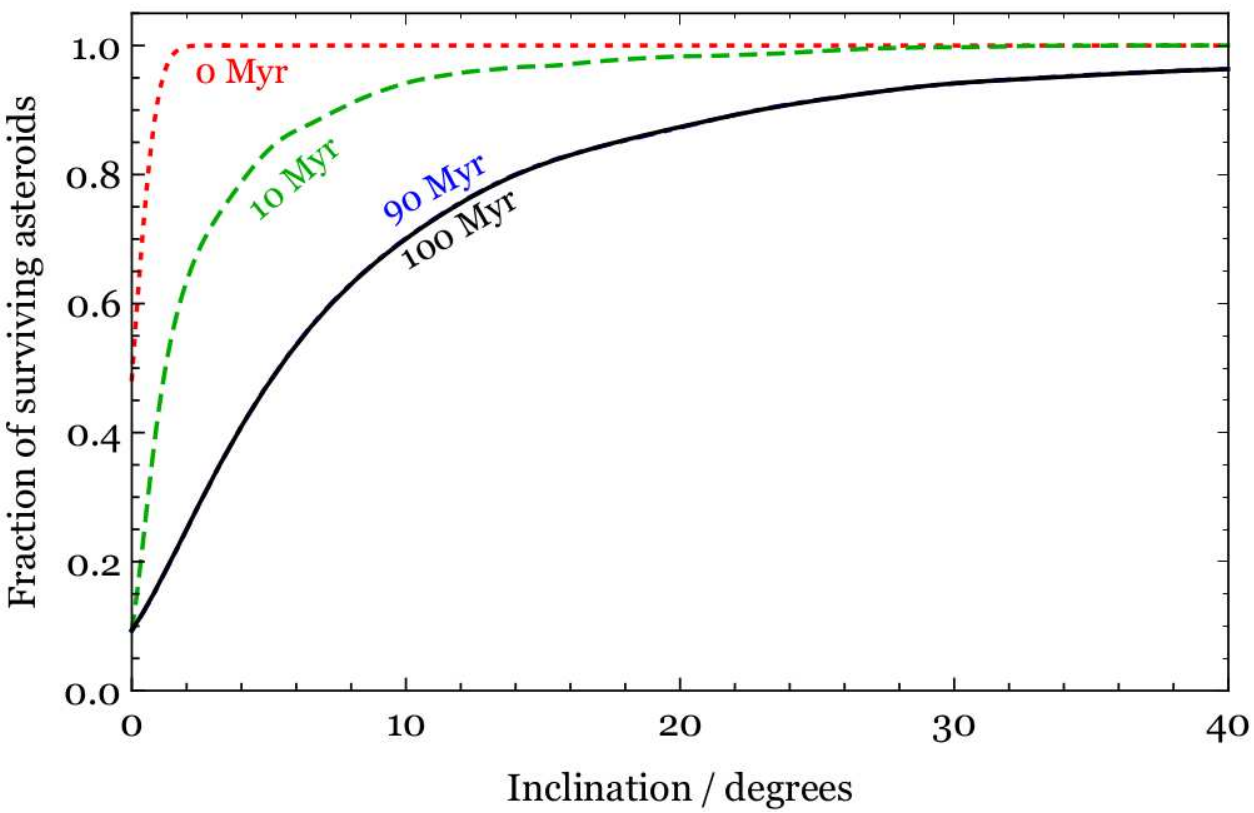}
\ \ \ \ \ \
\includegraphics[width=8.5cm]{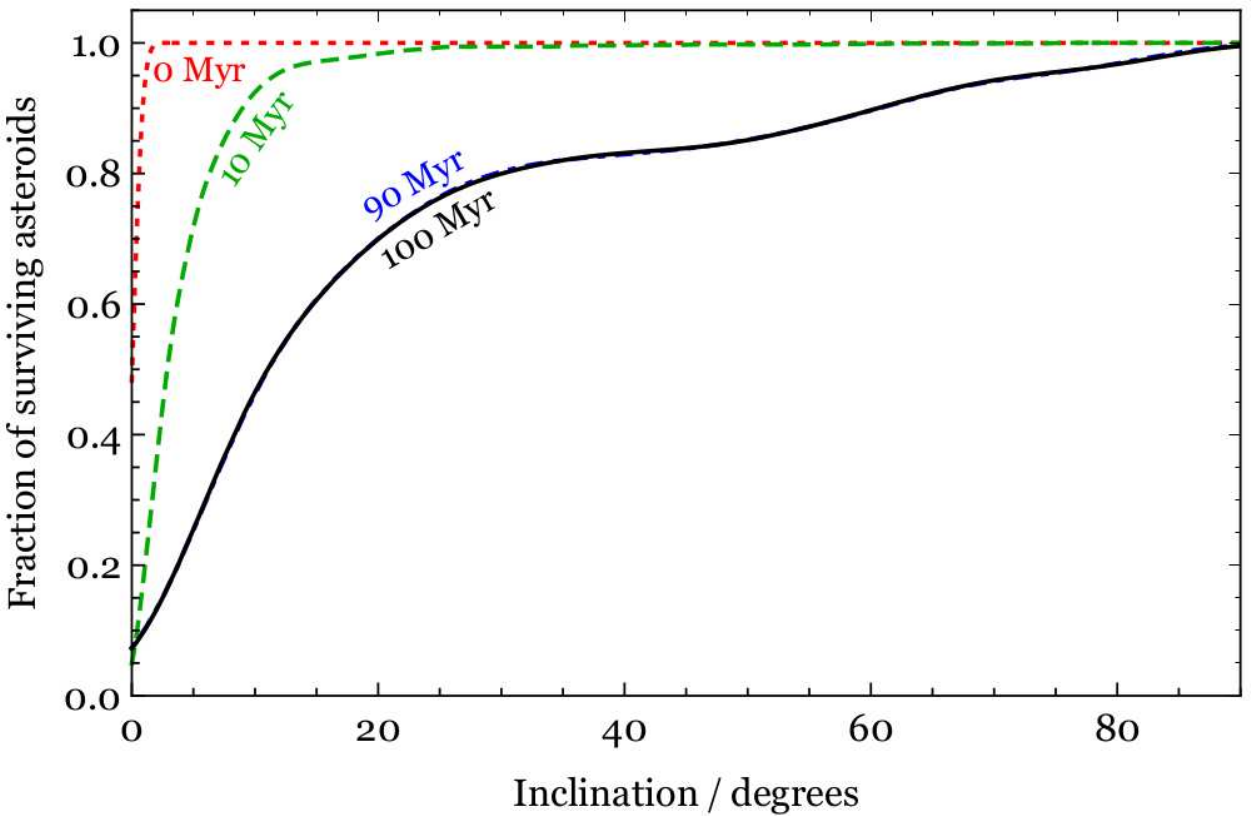}
}
\caption{
Cumulative distribution functions of semimajor axes (top panels), eccentricities (middle panels) and inclinations (bottom panels) for the time evolution of 2000 exo-planetesimals in initially $40-1000$~au orbits subject
to perturbations from a stellar birth cluster and four interior
giant planets. The left and right panels respectively correspond to two different systems. In all cases, the 90 Myr and 100 Myr curves
almost entirely overlap (the 90 Myr curves are dot-dashed blue curves while the 100 Myr curves are solid black), suggesting that by 100 Myr the systems have dynamically settled.
}
\label{Cluster1}
\end{figure*}
%%%%%%%%%%%%%%%% Figure (datasets 11 and 15)

Figure~\ref{Cluster1} provides detailed illustrations for two of these systems {\rev for the entire $40-1000$~au disc.} The left panels shows the system with 43 per cent surviving planetesimals (labelled ``system \#1'') and the right panels show the system with 9.7 per cent surviving planetesimals (labelled ``system \#2''). The top, middle and bottom panels respectively illustrate the semimajor axis, eccentricity
and inclination distributions. The left and right panels provide different viewpoints on each
distribution. {\rev The fraction of surviving planetesimals for the inner $40-150$~au regions of both systems is 100 per cent, and hence is not shown separately.}

The figure reveals that the outermost regions of each planetesimal disc are stripped. Other systems
(not shown) which retain nearly their entire planetesimal discs harbour a more homogeneous
distribution of semimajor axes throughout the 100 Myr cluster evolution. Figure \ref{Cluster1}
also illustrates how initially circular and co-planar planetesimals orbit distributions fan out in eccentricity
and inclination space. The four snapshots in time which are displayed in the figures demonstrate the
significant extent of the dynamical excitation from 0 Myr to 10 Myr and the insignificant evolution
from 90 Myr to 100 Myr; i.e. by 100 Myr, the systems have effectively dynamically settled (see also \citealt*{flybyetal2019}).

\section{Main-sequence evolution}

\label{sec:MS}

At $t=100$ Myr, we assume that the planetary systems have left the cluster environment and evolve for the remainder
of the star's main-sequence evolution ``in the field''. The duration of the main-sequence evolution (about 11 Gyr) is determined 
by assuming that the star's initial mass and metallicity are the same as the Sun's. Variations in this value will arise depending on the stellar model used, but do not noticeably affect the final qualitative result.

Unlike in the previous section, numerically integrating the planetary systems for the entire 11 Gyr is not computationally feasible.
The primary factors which restrict the timescale for such integrations are the total number of bodies and the presence of 
the exo-Jupiter (the planet with the shortest orbital period, unless a planetesimal is perturbed within 5 au of the star at some epoch). 
Therefore, we proceed [i] by arguing that the planetesimal evolution is negligible during this phase, and [ii] supporting this 
argument with limited duration integrations.

\subsection{Argument for negligible evolution}

We begin our argument by asserting that close stellar flybys in the field are infrequent and fast compared to those in the cluster environment. Further, the Milky Way Galaxy is a collisionless system. Because the relaxation time in the Solar neighborhood is longer than a Hubble time (see \citealt*{bintre2008}), major encounters are rare after the cluster dissolves \citep{2015MNRAS.451..144P}. Consequently, over an 11~Gyr timescale, the expected single closest encounter distance with a planetary system is on the order of several hundred au \citep{zaktre2004,vermoe2012,corgil2017}. Hence, for this particular closest encounter, whether an individual planetesimal at that distance would be significantly perturbed becomes a function of geometry. 

Additionally, because the main-sequence lifetime of 11 Gyr is two orders of magnitude higher than the 100 Myr cluster timescale,
additional {\rev processes} which act over long timescales need to be considered at this stage. Three of these {\rev processes} are Galactic tides, 
collisional grinding, and radiative forces.

\subsubsection{Galactic tides}

The consequences of Galactic tides are a strong function of stellar density (or location in the Milky Way),
inclination with respect to the Galactic disc, and planetesimal-star separation. If, for example, our simulated systems were in the Galactic
bulge, then Galactic tides acting over 10 Gyr would significantly affect planetesimal discs \citep{vereva2013a}. However,
in the Solar neighbourhood, the effect is muted: the orbital eccentricity of a highly-inclined Sedna-like object 
(with a semimajor axis of about 550 au and eccentricity of about 0.85) may change by a maximum of about $0.05$ 
due to tides \citep{vereva2013b}. Because {\rev all of our planetesimals have semimajor axes within a factor of two of Sedna's}, their eccentricity changes due to tides are likely
to be comparable to those from stellar flybys (just over much longer timescales). 

We note that the Galactic environment of a planetary system likely changes with time. In fact, stars migrate in the Galaxy through interaction with time-varying spiral arms in a process called ``churning'' \citep{SellwoodBinney2002}. Our present understanding of the Galactic metallicity gradients and age--metallicity relations suggests that the Sun was born 1--3\,kpc closer to the Galactic centre than its current location \citep{Minchev+13,Minchev+18,Frankel+18,Feltzing+19}. This distance is comparable to the scale length of the Galactic thin disc \citep{BlandHawthornGerhard16}, and so field star densities at the Solar birth radius could have been higher than the present value by a factor of a few. We are therefore slightly underestimating the impact of field star encounters and Galactic tides in this section.

\subsubsection{Collisional grinding}

Collisional grinding amongst planetesimals would not eject
planetesimals nor their fragments from their original annulus. {\rev Instead,} the consequence of mutual collisions between 
planetesimals is a change in their size distribution 
\citep{dohnanyi1969,botetal2005}. This alteration does not affect {\rev the way we represent our systems in our numerical integrations} because we model our planetesimals as massless point particles, {\rev and do not make assumptions about the initial masses of our disc. Further, the physical evolution of our surviving large planetesimals ($> 100$ km) is independent of our main results, because we know {\it a posteriori} that these objects will not pollute the eventual white dwarf, and become interstellar planetesimals at a negligible rate compared with those generated from the initial cluster evolution.}

{\rev The consequence of changing the size distribution is that the ground-down planetesimals may become small enough to be affected physically and orbitally by stellar radiation (see Section 3.1.3), particularly during the giant branch phases of stellar evolution (see Section 4.1.3). Hence, if one imposes a size and mass distribution onto our massless particles, then depending on these parameters, no 100 km-sized planetesimals may survive throughout the main sequence. The actual collisional lifetimes of objects are a nontrivial function of many parameters, including the dispersal threshold for collisions (critical specific energy), the eccentricity distribution, the inclination distribution, and the breaking radius between the gravity and strength-dominated regimes \citep[e.g.][]{wyaetal2007,lohetal2008}.  For typical parameters assumed or calculated for extrasolar debris disks, collisional evolution is not significant for objects above 100 km in size \citep[e.g.,][]{keny2008,koba2014,krietal2018}.  So, in this paper, we consider only surviving planetesimals larger than 100 km (which do not {\revv change their orbital elements} due to collisions)\footnote{
{\rev For added perspective, in the solar system,} the collisional evolution of the Main Belt asteroids has been investigated in a high-level
of detail within subsets of the belt itself \citep[e.g.][]{cibetal2014}. The collisional evolution of the Kuiper Belt
is more speculative, and often based on the fraction of binary asteroids \citep{nesetal2011,deletal2012,bruzan2016}.
Nevertheless, Kuiper Belt objects greater than about 10 km in size are not thought to have undergone
collisional evolution over the last 4 Gyr or so \citep{bramor2013,jutetal2017,mornes2019} and hence represent 
remnants of early dynamical instability
which involved Neptune \citep{woletal2012,parker2015,cheetal2016,nesvok2016,volmal2019}. {\rev However, Sedna may be a remnant of collisional evolution far beyond the Kuiper belt \citep{siltre2018}.}}.
}

\subsubsection{Radiative forces}

Important radiative forces on planetesimals primarily arise from the YORP and Yarkovsky effects \citep{voketal2015}. 
These forces describe the orbital movement (Yarkovsky effect) and the spin changes (YORP effect) from thermal imbalances created by nonzero thermal inertia and anisotropically emitted thermal radiation. The secular consequences of both effects have been observationally verified within the Main Belt, despite the fantastically
small accelerations produced by the Sun's radiation (often on the order of 1 pm s$^{-2}$ for the Yarkovsky effect). The YORP effect can break apart planetesimals by spinning them up to the point of rotational fission \citep{hols2007,warnetal2009,poletal2017}, which primarily affects the size distribution of the fragments, rather than their orbits. Both the Yarkovsky and YORP effects are negligible at distances of at least 40~au along the main sequence, and so their contribution here can be ignored.

Overall, we argue that planetesimal disc objects with sizes $\gtrsim 100$ km have orbits which vary negligibly 
during the 11 Gyr main-sequence phase of stellar evolution. The extent of their orbital variation should be 
eccentricity shifts on the order of hundredths, unless a stellar flyby achieves a particularly close encounter 
within hundreds of astronomical units.

\subsection{Short simulations}

We now attempt to support our argument by conducting feasibly short (hundreds of Myr) $N$-body simulations 
with a limited number of planetesimals (200 per simulation) but including both Galactic tides and stellar flybys. 
The results of these simulations may then be extrapolated over the entire main-sequence. The planetesimals are
treated as test particles and hence are not assigned masses.

We perform these simulations using the RADAU (variable timestep) integrator in the \textsc{Mercury} integration package \citep{chambers1999}
with implemented routines for Galactic tides from \cite{vereva2013a} and stellar flybys from \cite{veretal2014b}. We
assume our systems reside in the Solar neighbourhood and adopt the corresponding numerical values associated with tides
and flybys from those investigations: for the flybys, we assume a spatial stellar density of 0.392~pc$^{-3}$ \citep{paretal2011} and an encounter velocity of 46~km/s \citep{garetal2001}.
We also assume that the input values of the orbital elements of the planetesimals and giant planets equal the corresponding
output values from our cluster simulations\footnote{{\rev In our stellar cluster simulations in Section~2, the cluster evolution does slightly change the orbits of the giant planets, with the exo-Neptune experiencing the largest change. The typical scale of the changes for the exo-Neptune are 0.01-0.1 au in semimajor axis, 0.005-0.025 in eccentricity, and $0.001^{\circ}-1.0^{\circ}$ in inclination.}}.

%%%%%%%%%%%%%%%% Figure (dataset 11)
\begin{figure*}
\centerline{\Large \bf Main-Sequence Evolution}
\centerline{}
{\Large\bf System \#1 with four giant planets \ \ \ \ \ \ \ \ \ \ \ \ System \#1 without four giant planets}
\centerline{
\includegraphics[width=8.5cm]{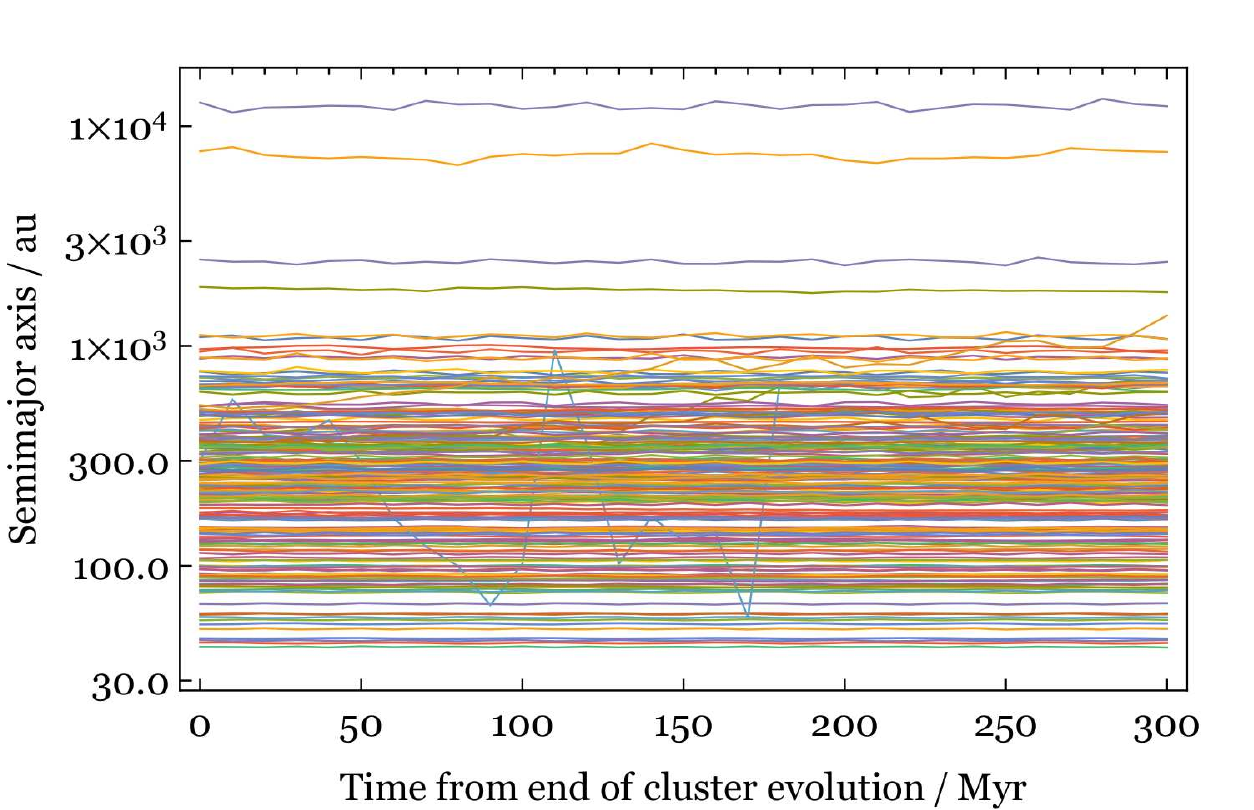}
\ \ \ \ \ \
\includegraphics[width=8.5cm]{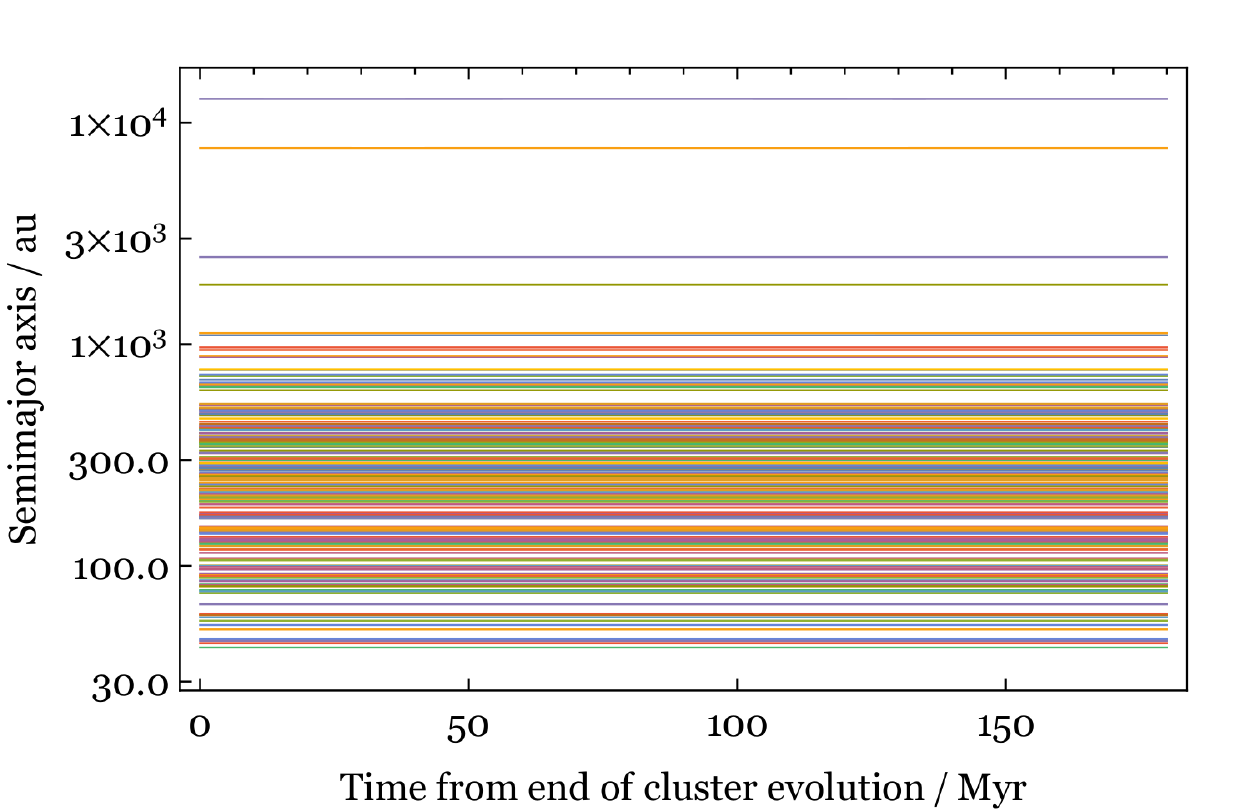}
}
\centerline{}
\centerline{
\includegraphics[width=8.5cm]{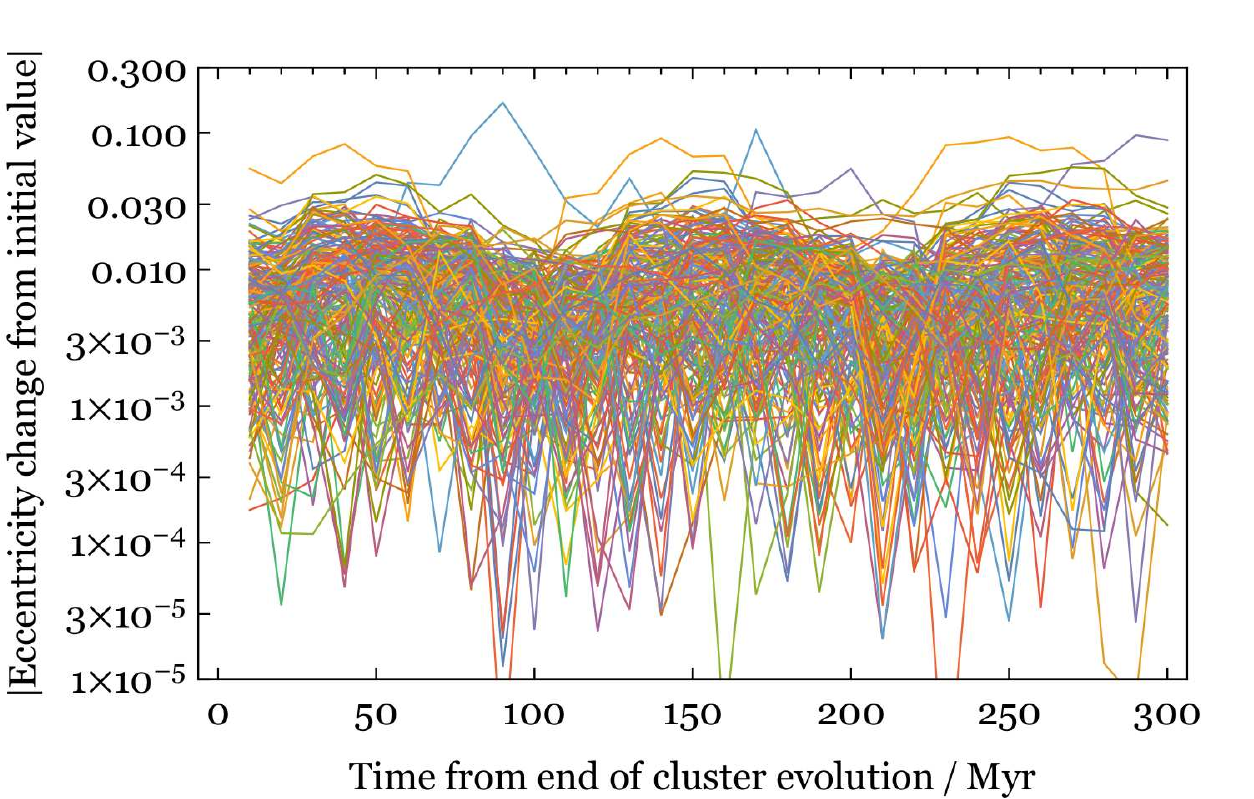}
\ \ \ \ \ \
\includegraphics[width=8.5cm]{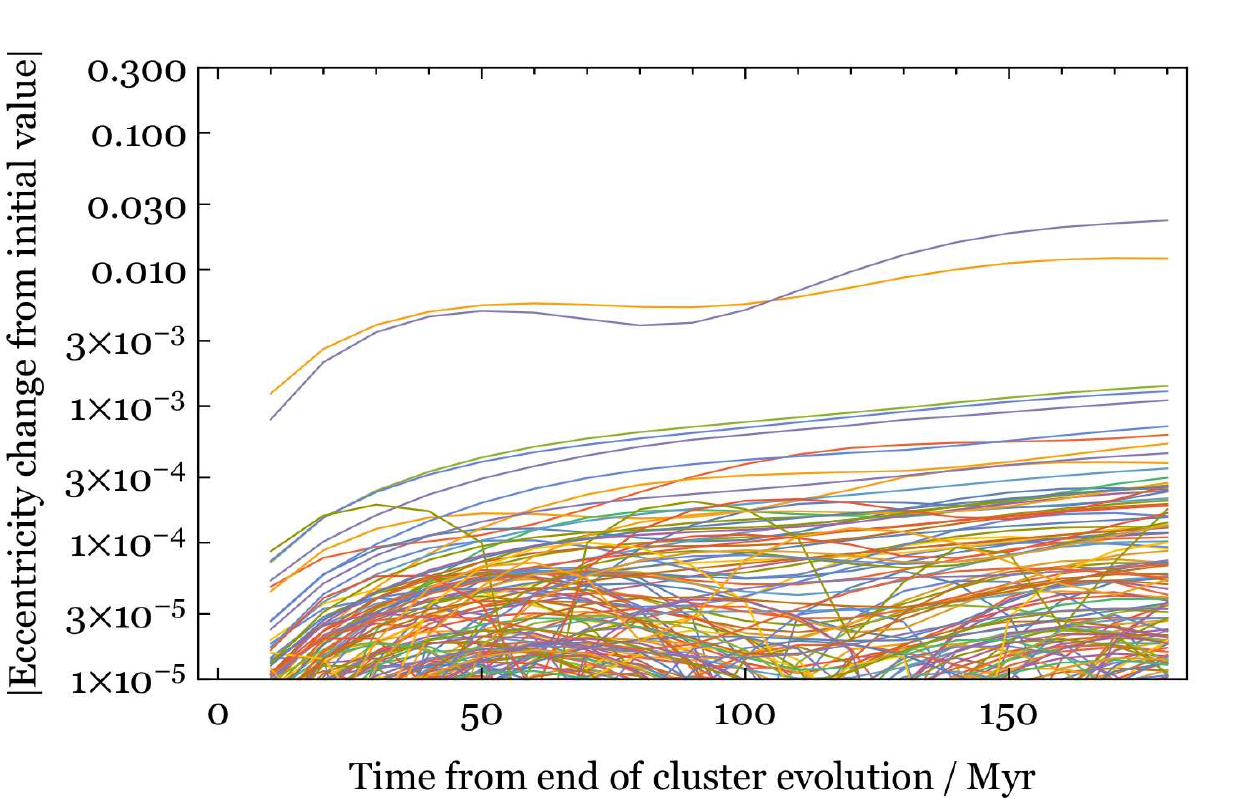}
}
\centerline{}
\centerline{
\includegraphics[width=8.5cm]{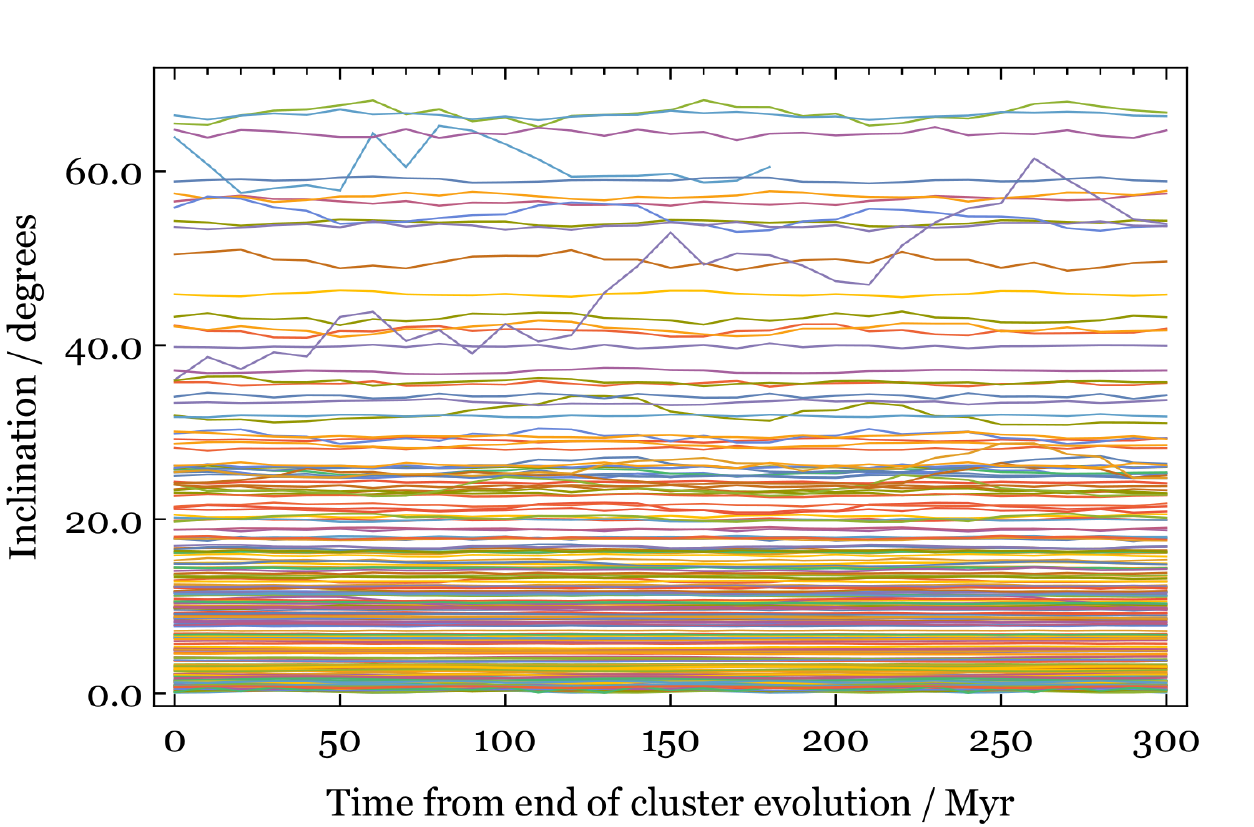}
\ \ \ \ \ \
\includegraphics[width=8.5cm]{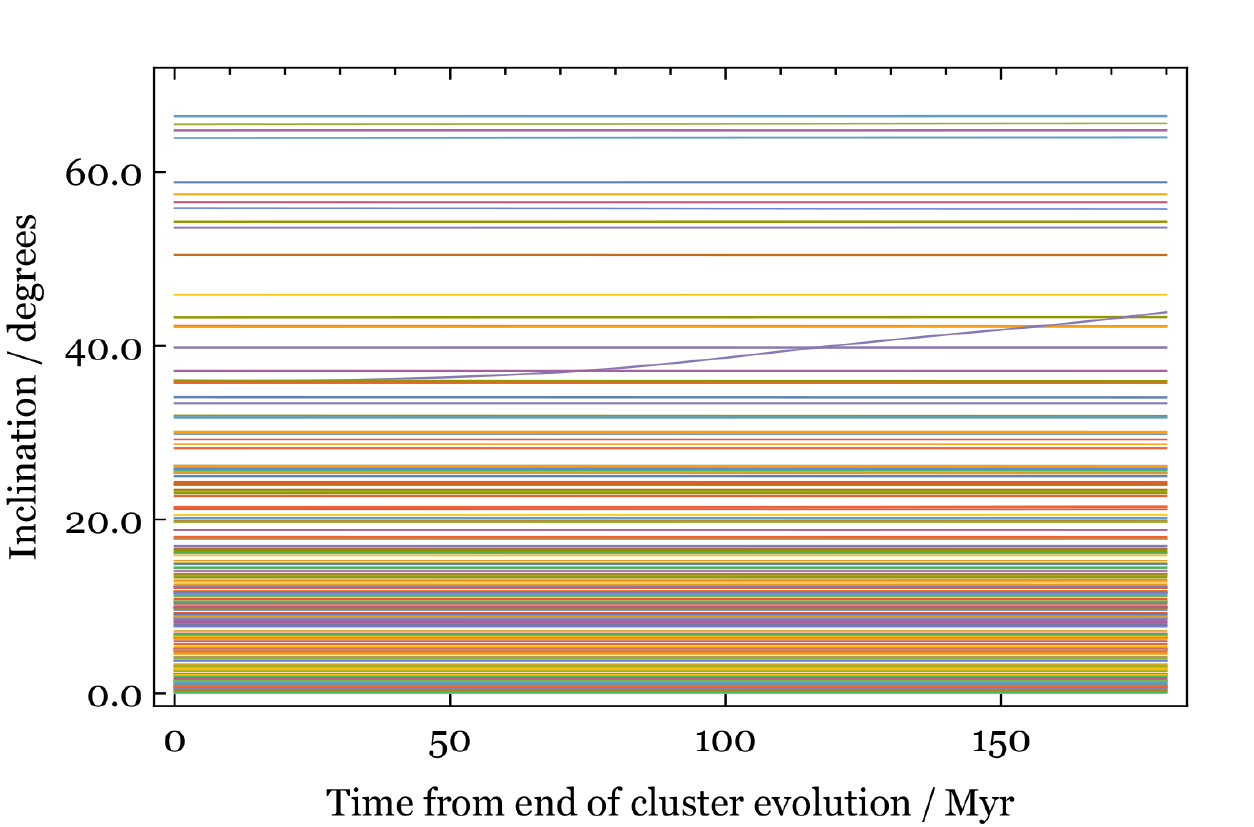}
}
\caption{
Main-sequence evolution (for hundreds of Myr only) 
of 200 planetesimals starting
with their post-cluster evolution orbital parameters
from Fig. \ref{Cluster1}.
Included on all panels are the effects of stellar flybys
and Galactic tides. The left panels also include the
effect of an exo-Jupiter, exo-Saturn, exo-Uranus and
exo-Neptune. The plots illustrate that the major planets
are the strongest perturbers of exo-planetesimals, and that
despite their presence, planetesimal discs largely
maintain their structure in the absence of a violent
dynamical instability.
}
\label{MS1}
\end{figure*}
%%%%%%%%%%%%%%%% Figure (dataset 11)

%%%%%%%%%%%%%%%% Figure (dataset 122)
\begin{figure}
\centerline{\Large \bf Main-Sequence Evolution}
\centerline{}
\centerline{\Large\bf System \#3 with four giant planets}
\centerline{
\includegraphics[width=8.5cm]{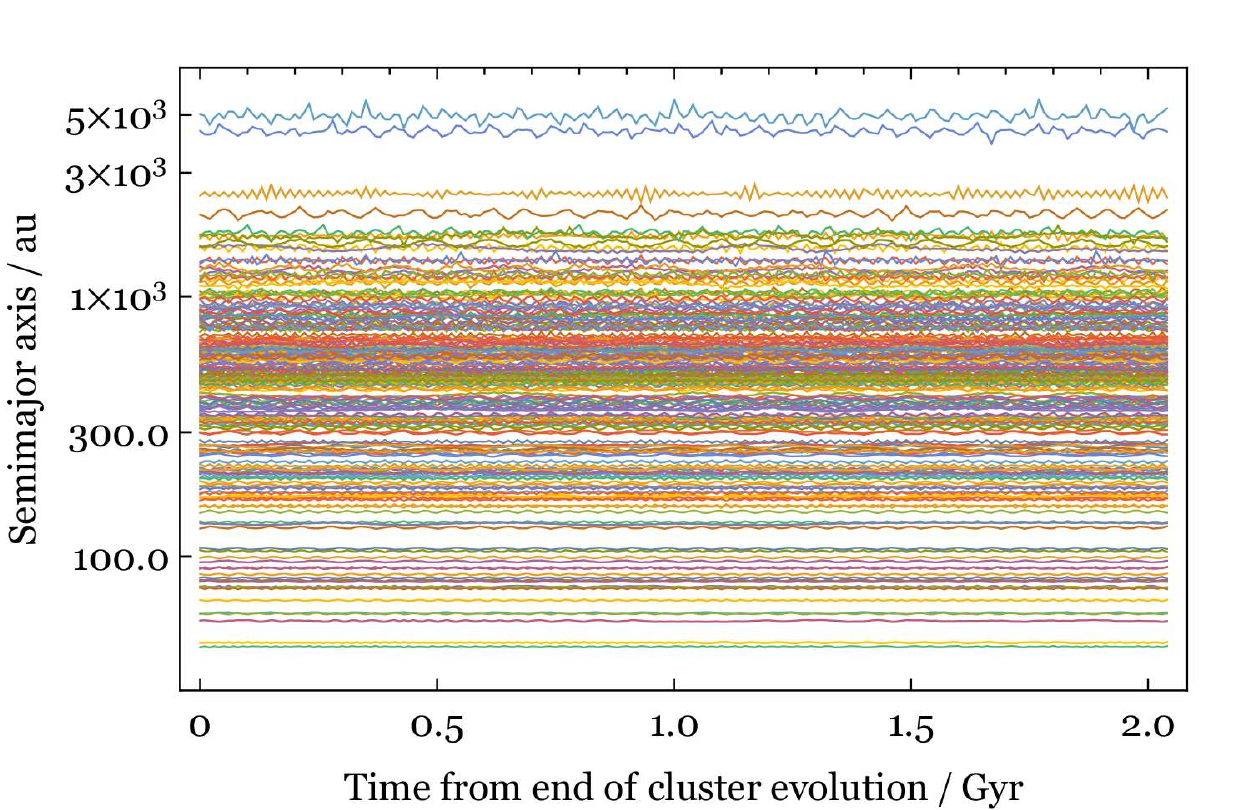}
}
\centerline{}
\centerline{
\includegraphics[width=8.5cm]{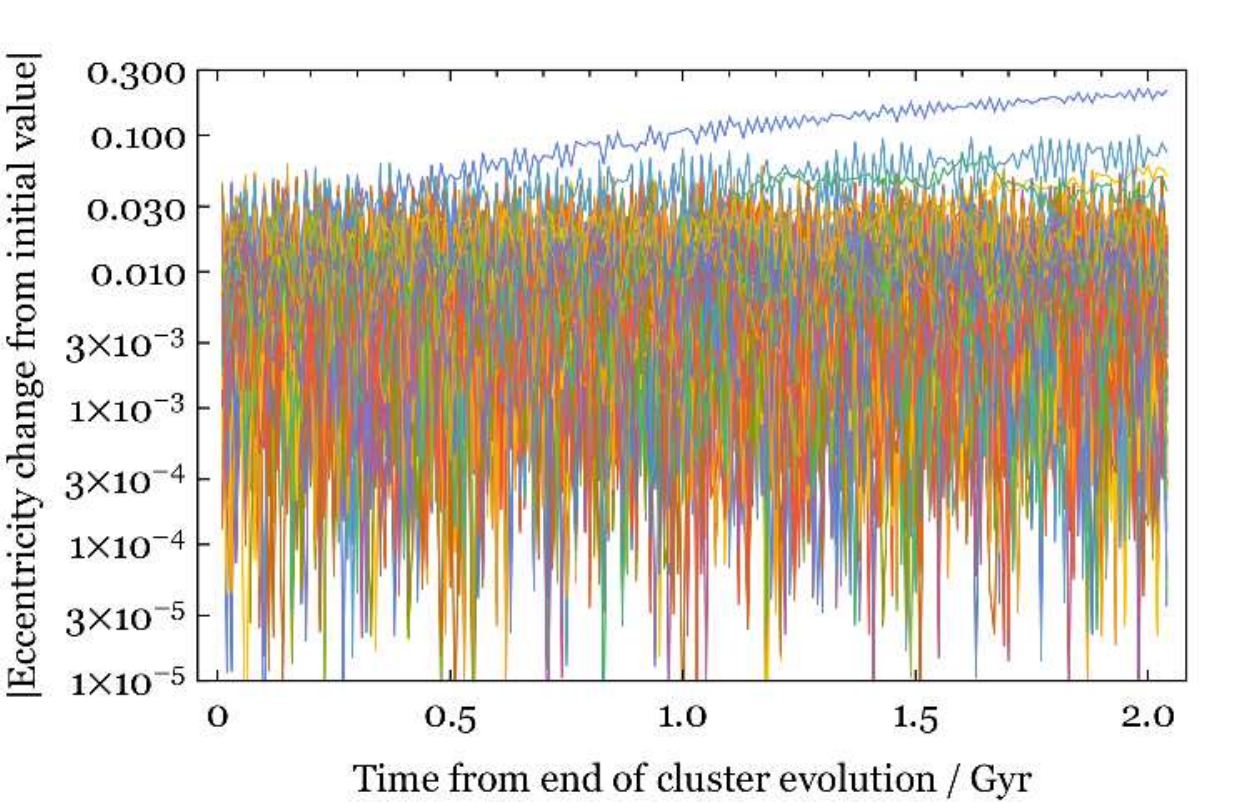}
}
\centerline{}
\centerline{
\includegraphics[width=8.5cm]{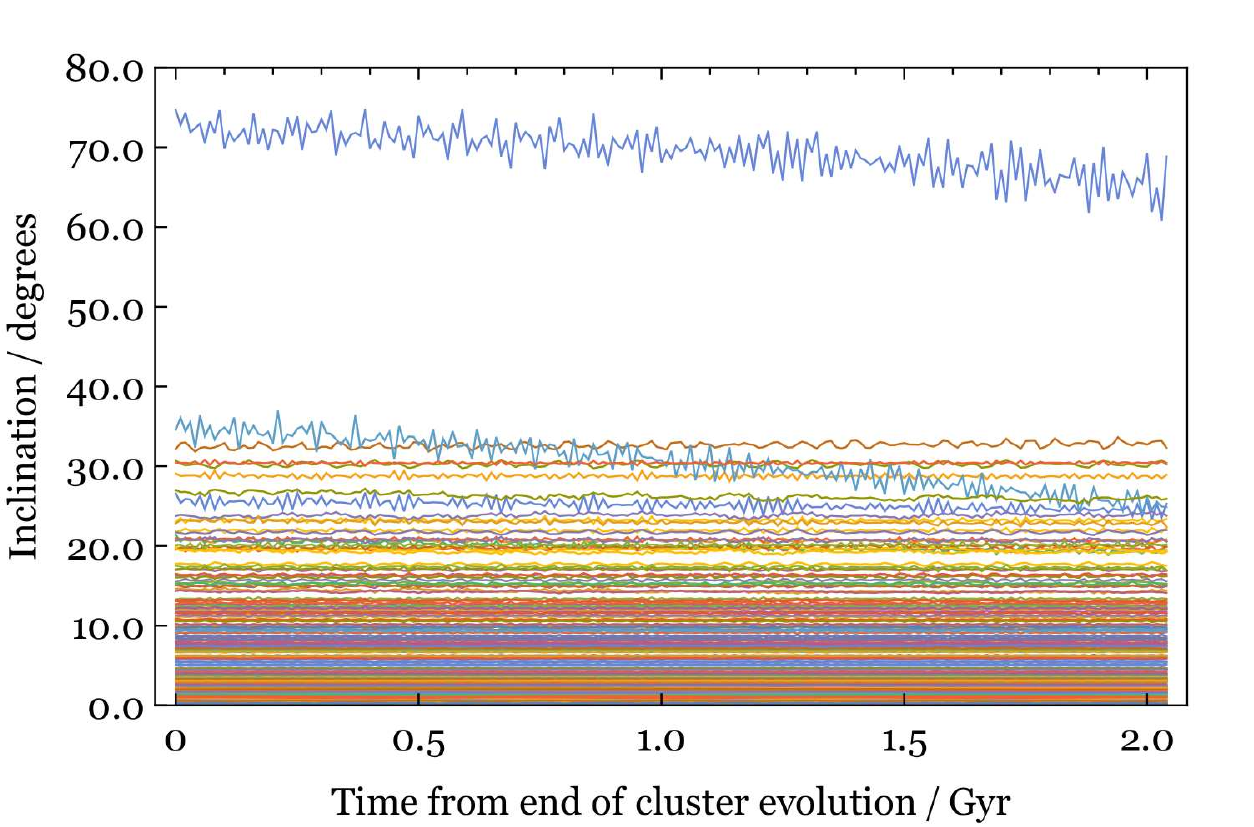}
}
\caption{
Like Fig.~\ref{MS1}, but for a different system where the cluster
evolution (not shown in Fig. \ref{Cluster1}) 
was ineffectual at removing planetesimals from an initial
Kuiper belt (such that 98 per cent of the planetesimals remained stable) and
for a longer simulation timescale (2 Gyr). The resulting trends
are similar to those in Fig. \ref{MS1}. The two planetesimals with the
greatest separations and initial inclinations showcase discernable
secular trends in their eccentricity and inclination temporal evolution
profiles due to Galactic tides.
}
\label{MS2}
\end{figure}
%%%%%%%%%%%%%%%% Figure (dataset 122)

Figure \ref{MS1} illustrates the time evolution of 200 {\rev randomly chosen} planetesimals which survived
the cluster evolution from the left panels of Fig. \ref{Cluster1}. The left panels of Figure \ref{MS1} include the four giant
planets, and the right panels do not, in order to illustrate that these planets represent
the greatest driver of planetesimal orbital evolution. The middle panels display the eccentricity
change from the initial values, highlighting how major planets usually {\rev vary} planetesimal eccentricities {\rev either increasing or decreasing them}
by orders of magnitude more than stellar flybys or Galactic tides. In the simulations in the
left panel, the closest stellar encounter occurred at a distance of 1786 au at 226 Myr into
the simulation, whereas for the simulations in the right panel, the closest stellar encounter
was at 7842 au at 9.5 Myr into the simulation. Note that around these times there is no
discernibly significant change in the orbital parameters.

In all cases, the eccentricity
variation rarely exceeds 0.03. The two highest curves
on the middle right panel correspond to the two planetesimals with semimajor axes of about $10^4$ au
and illustrate the secular oscillations produced from the Galactic tide; these planetesimals
are outliers to the main distribution. In the simulations with major planets, only one of 
these 200 planetesimals becomes unstable (the jagged blue line in the upper left panel) and 
features the greatest eccentricity change of any planetesimal\footnote{We note that if some planetesimal disc objects were initially highly eccentric, then these relatively modest eccentricity changes would be much more significant. For example, consider a body at 1000\,au with a pericentre at 40\,au and an eccentricity of 0.96. If its eccentricity were increased by 0.03, its pericentre would now be at just 10\,au, close enough to interact with Saturn.}.

Figure \ref{MS2} displays results for a different system (denoted ``System \#3'') which
retained 98 per cent of its planetesimals after cluster evolution, and for which we ran the 
integration (200 planetesimals only) for much longer (2 Gyr) than System \#1. Despite the
longer timescale, the overall result is the same as Fig. \ref{MS1}: planetesimal discs maintain
their post-cluster orbital elements except for the outliers. The longer timescale also
allows oscillations in the orbital element distributions of these outliers (the furthest planetesimals)
to be more readily detected. These oscillations arise from a combination
of stellar flybys and Galactic tides. The amplitude of the inclination oscillations is highest
for the most inclined (relative to the Galactic plane) planetesimal. Stellar flybys again do not
appear to significantly affect the planetesimal dynamics, despite 35 close approaches within $10^4$ au, with
the closest approach located at 1237 au at 1.52 Gyr into the simulation.

\section{Giant branch evolution}
\label{sec:postMS}

{\rev In our short main sequence simulations,} the orbital structure of the planetesimal discs remained effectively static. {\rev By assuming that later in the main sequence the disc was not significantly disturbed by a gravitational instability nor a close stellar flyby, we now consider the consequences of giant branch evolution.} Giant stars undergo significant physical changes: a $1.0M_{\odot}$ star will inflate its envelope
out to a distance of about 1 au, lose about half of its mass through stellar winds, and increase its luminosity by a factor 
of about $4 \times 10^3$.

\subsection{Description of effects}

\subsubsection{Stellar engulfment}

The expansion of the stellar envelope can directly engulf closely orbiting planets, and tidally draw into the envelope
planets which reside beyond the maximum extent of the stellar envelope. This critical engulfment distance has
been extensively investigated, but nevertheless varies depending on the stellar and tidal models adopted
\citep{kunetal2011,musvil2012,adablo2013,norspi2013,viletal2014,madetal2016,staetal2016,galetal2017,raoetal2018,sunetal2018}. 
Regardless, all models agree that objects orbiting a $1.0M_{\odot}$ star beyond a few au would avoid engulfment.

Although planetesimal discs would survive engulfment, they could be affected by tides indirectly. Engulfment of
planets changes the secular resonance structure of the system \citep{petmun2017,smaetal2018,smaetal2019},
which may trigger instabilities (collisions or ejections) in previously stable regions. However, for solar system analogues 
\citep{smaetal2018,smaetal2019}, where an exo-Mercury, exo-Venus and maybe an exo-Earth 
would be engulfed \citep{schcon2008}, the secular resonance region of greatest importance 
would be internal to the giant planets. Hence, we need not consider engulfment for our planetesimal discs.

\subsubsection{Stellar mass loss}

Stellar mass loss, however, must be considered. The changing potential expands the orbits of all planets
and planetesimals \citep{omarov1962,hadjidemetriou1963}. If the mass loss is assumed to be isotropic
\citep{veretal2013a}, then objects within about $10^3$ au (including the planetesimal discs and four
giant planets) would all double their semimajor axis. The extent to which their eccentricities change
is positively correlated with semimajor axis, and usually neglected below a certain ``adiabatic'' limit
\citep{veretal2011}. For distances of $10^3$ au, we should expect eccentricity shifts just on the order of hundredths.
Hence, the result of stellar mass loss on our planetesimal discs would be a larger, but self-similar, annulus.

However, although the mutual semimajor axis ratios between planetesimals and planets would not change
due to adiabatic mass loss,
the decrease in mass from the central star may nevertheless trigger instability \citep{debsig2002}. Whether and when this
trigger is activated depends on number of bodies, mass of bodies and mutual separation between
the bodies 
\citep{bonetal2011,debetal2012,musetal2013,portegieszwart2013,veretal2013b,frehan2014,musetal2014,vergae2015,veretal2016b,veretal2017b,veretal2018}. 
We will explore the extent of the potential instability in our planetesimal discs 
due to the presence of an exo-Neptune through numerical simulations. 

\subsubsection{Stellar luminosity}

First, we comment on the effects of the increased luminosity of the host star during the giant branch phases. While grain-sized particles may be ``blown-out" by {\revv radiation pressure \citep{bonwya2010,donetal2010,maretal2020,zotver2020}}, larger minor planets are subject to other effects. For example, the enhanced YORP effect would effectively destroy solar system Main Belt asteroids between 100 m and 10 km in size due to rotational spin-up \citep{veretal2014a}. Because the planetesimal's change in spin rate is inversely 
proportional to the square of both the separation and planetesimal radius, the YORP effect on 100 km planetesimals would
be at least four orders of magnitude smaller. Hence, we can safely neglect the YORP effect in post-main-sequence planetesimal
discs with planetesimals larger than about 100 km.

However, planetesimals which survive YORP-induced spin-up may be orbitally perturbed by the enhanced Yarkovsky effect.
We can estimate the extent of the Yarkovsky effect during the giant branch phase for a $1.0M_{\odot}$ main-sequence star
by using equations 108 and 110 of \cite{veretal2015b}.
Across the tip of the asymptotic giant branch, where the stellar luminosity is greatest, a planetesimal with a radius of 
100~km and a distance of 40~au would shift\footnote{These estimates are about the same orders of magnitude as the excitations expected from the integrated monotonic Yarkovsky drift across the 11 Gyr main-sequence evolution.} its semimajor axis by about $10^{-3}$ au and its eccentricity
by about $10^{-5}$. A shift of $10^{-3}$~au is negligible because that value is orders of magnitude smaller 
than the typical libration widths of strong resonances in the trans-Neptunian region. \cite{veretal2019} provide 
further evidence that the Yarkovsky effect for 100 km planetesimals in this region is negligible by considering
a variety of limiting Yarkovsky models that place bounds on the motion. 

\subsubsection{Galactic tides and stellar flybys}

Regarding Galactic tides and stellar flybys during the giant branch phases, the primary difference from
the main-sequence phase is the timescale over which these perturbations act. The durations of the 
main-sequence, red giant branch and asymptotic giant branch phases for a Sun-like star
are about 11 Gyr, 1.5 Gyr and 5 Myr, respectively. These differences indicate that Galactic tides 
and stellar flybys are less disruptive during the giant branch phases than on the main sequence. 
Figure 3 of \cite{veretal2014c}
plots the relative importance of Galactic tides, stellar flybys and stellar mass loss for the
planets orbiting red giant branch and asymptotic giant branch stars as a function of stellar 
mass and Galactocentric distance. That figure illustrates that we can neglect Galactic tides and 
stellar flybys during the giant branch phases of stellar evolution.

\subsubsection{Collisional grinding}

{\rev Collisional evolution within the disc during giant branch evolution follows the description in Section 3.1.2, except now the stellar mass is a function of time. \cite{bonwya2010} analyzed this case, and found that the collisional lifetime increases as the star loses mass. The primary reason is because this stellar mass loss expands the disc and increases the pairwise distance amongst all of its contents. The planetesimals are hence ``safer" during giant branch evolution than along the main sequence.
}

\subsection{Numerical simulations}

We now perform $N$-body numerical simulations to quantify the changes in the orbital architectures 
of planetesimal discs during the giant branch phases of evolution. The above discussion suggests that we can 
neglect radiative forces (assuming our test particles are larger than about 100 km, {\rev which is our assumption}), any residual effects 
from the engulfment of terrestrial planets, and external perturbers.
What remains to be explored by these simulations is potential instability triggered by mass loss with Neptune
and the planetesimals (we know that the giant planets themselves will remain stable from \citealt*{veras2016b}).

The code we adopt in this section is different from the individual codes used in Sections 2 and 3 (see Fig. \ref{fig:cartoon}). Here, because we must concurrently
model stellar and planetary evolution, we use the code presented in \cite{musetal2018}. This code is an updated
version from the one that was first presented in \cite{veretal2013b} and uses the RADAU integrator from the
\textsc{Mercury} integration package \citep{chambers1999}. 

We choose RADAU as the integrator because it is non-symplectic (allowing it to handle an arbitrarily-changing potential) but very accurate, being able to accurately track not only a planet's semimajor axis but also its orbital phase. We adopt a tolerance value for the RADAU integrator of $10^{-11}$. The stellar evolution input is provided by the {\tt SSE} code \citep{huretal2000}. The output from the stellar evolution code is fed to \textsc{Mercury,} where the stellar mass is updated at every subdivision of a timestep for the force calculations. The stellar radius is updated every timestep, when checks are made for the removal of particles by collision with the star (no collisions occurred in our simulations).

We adopt the default numerical parameters in the {\tt SSE} code for our simulations. Among these is the Reimers mass
loss coefficient, which is set to 0.5.  This coefficient dictates the time evolution of the mass loss. \cite{verwya2012}
sampled a realistic range of these coefficients for the future evolution of the Sun, and found that the total amount
of mass lost ranges from about $0.465M_{\odot}$ to $0.490M_{\odot}$ depending on the choice of coefficient.
Also, the time at which the asymptotic giant branch phase is initiated can vary by hundreds of Myr depending on this coefficient
choice. Further, higher coefficients produce greater mass loss rates, yielding a solar system post-main-sequence escape boundary
range of $10^3 - 10^4$ au (well beyond the outer boundary of most of our planetesimals). These variations due to 
coefficient choice are not sufficiently large to warrant partitioning the available computational resources in 
order to sample different coefficients.

As justified in Section 3, we adopt for our initial conditions the outputted values of the simulations performed in Section 2.
We start our simulations right before the start of the red giant branch phase, which is 10.94 Gyr after the instance of 
Zero-Age-Main-Sequence (ZAMS). We run these simulations for the entire duration of the giant branch phases (1.5 Gyr) 
plus during the start of the white dwarf phase, for 8 Gyr in System \#1 and 2.4 Gyr in another system denoted System \#4.

%%%%%%%%%%%%%%%% Figure (datasets 11 and 15)
\begin{figure}
\centerline{\Large\bf Post-Main-Sequence Evolution: System \#1}
\centerline{
\includegraphics[width=8.5cm]{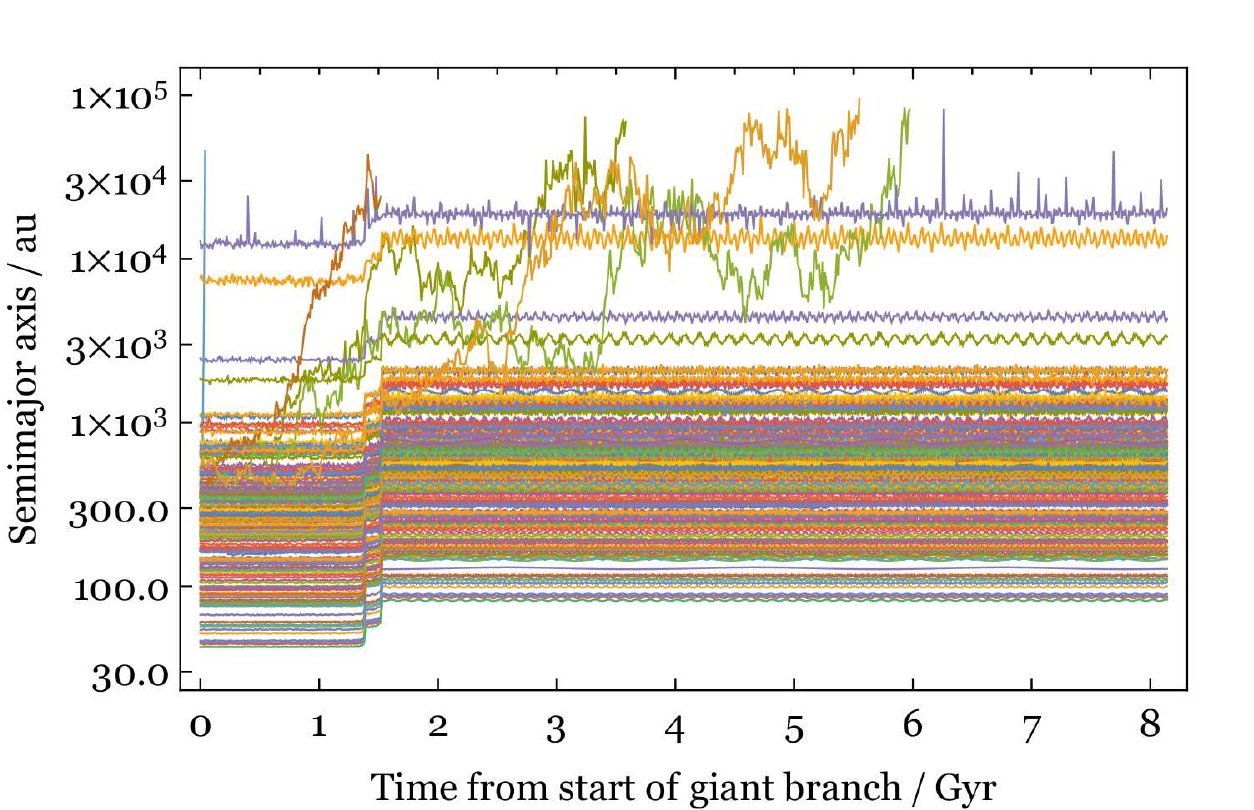}
}
\centerline{}
\centerline{
\includegraphics[width=8.5cm]{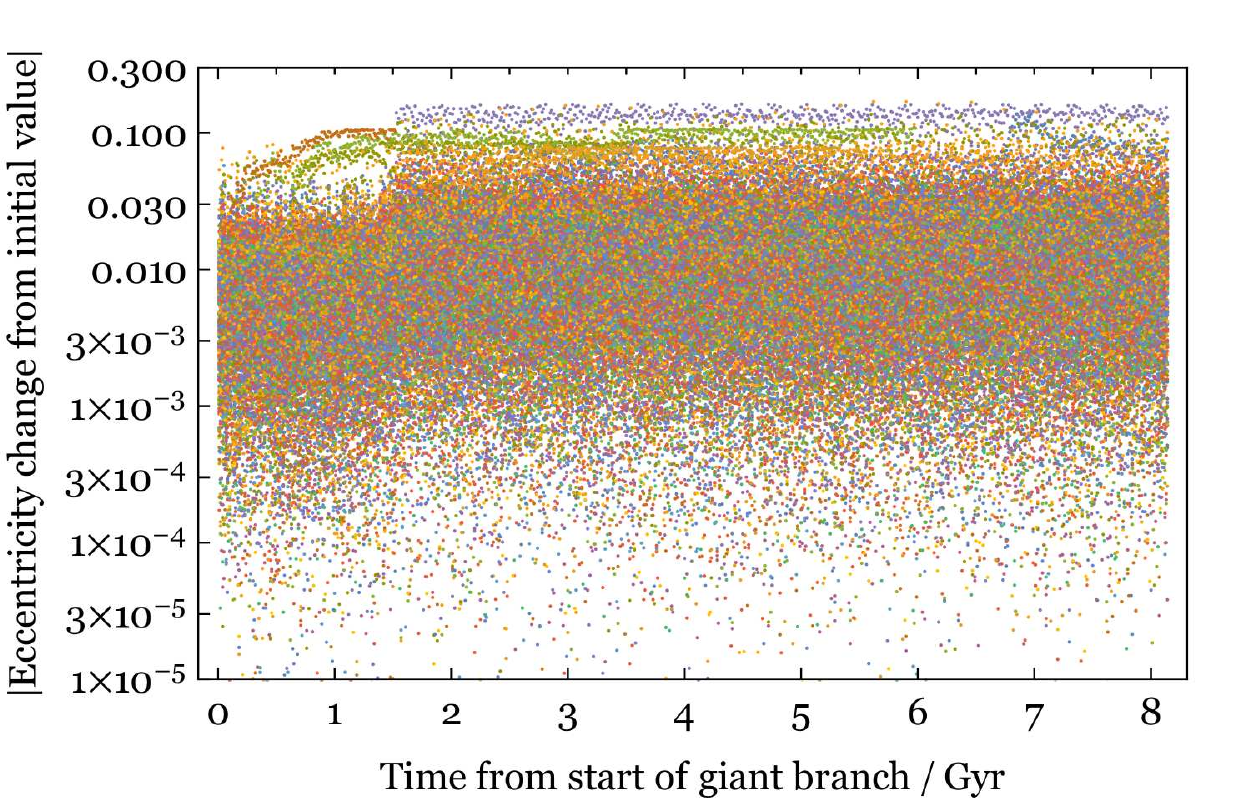}
}
\centerline{}
\centerline{
\includegraphics[width=8.5cm]{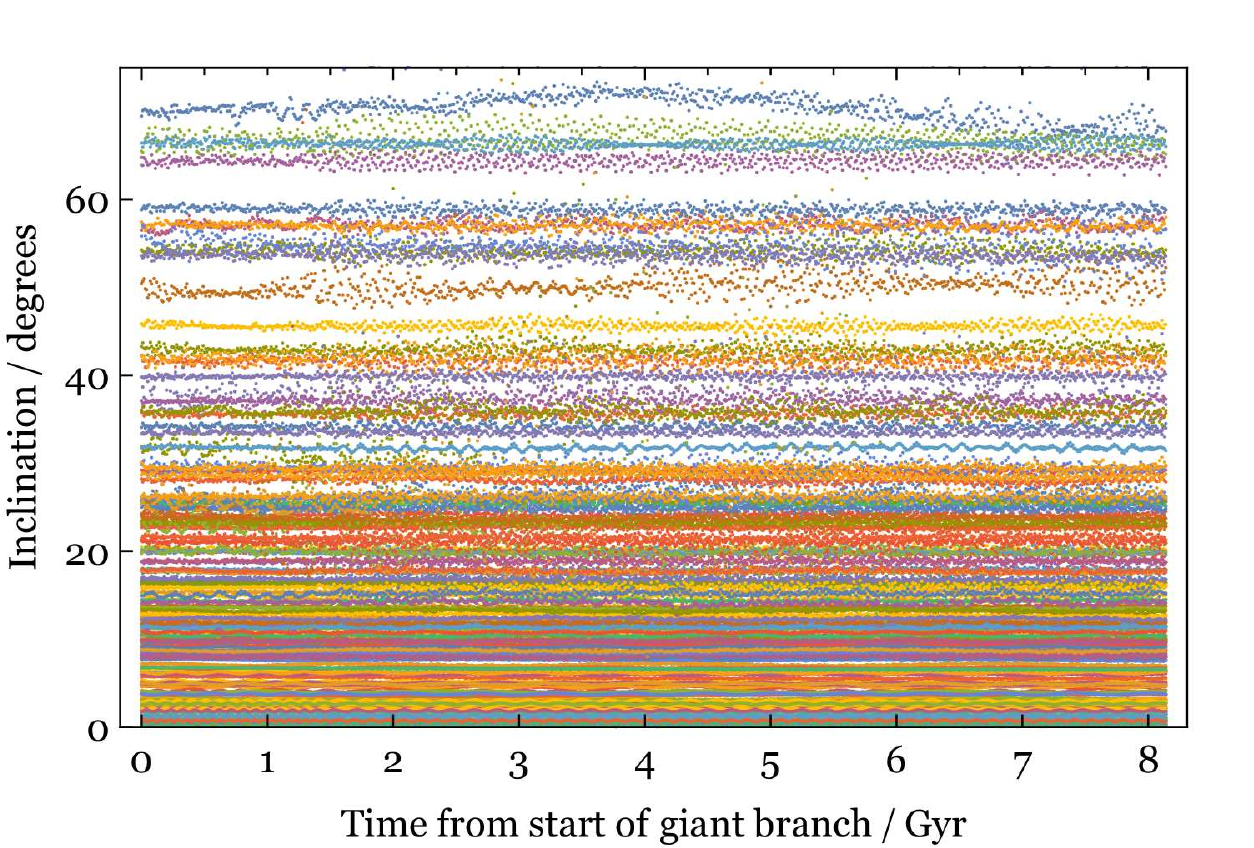}
}
\caption{
Post-main-sequence evolution for the same system in
the left panels of Fig. \ref{Cluster1} and in Fig. \ref{MS1}. The evolutions in the
bottom two panels are illustrated with points rather than
joined lines for greater clarity. Despite a few instabilities,
97.5 per cent of the post-cluster planetesimal disc remains
intact, with an eccentricity variation of typically no more than
a few hundredths, and an expected doubling of its semimajor 
axis.
}
\label{postMS1}
\end{figure}
%%%%%%%%%%%%%%%% Figure (dataset 11 and 15)

%%%%%%%%%%%%%%%% Figure (dataset 110)
\begin{figure*}
\centerline{\Large\bf Post-Main-Sequence Evolution: System \#4}
\centerline{
\includegraphics[width=8.5cm]{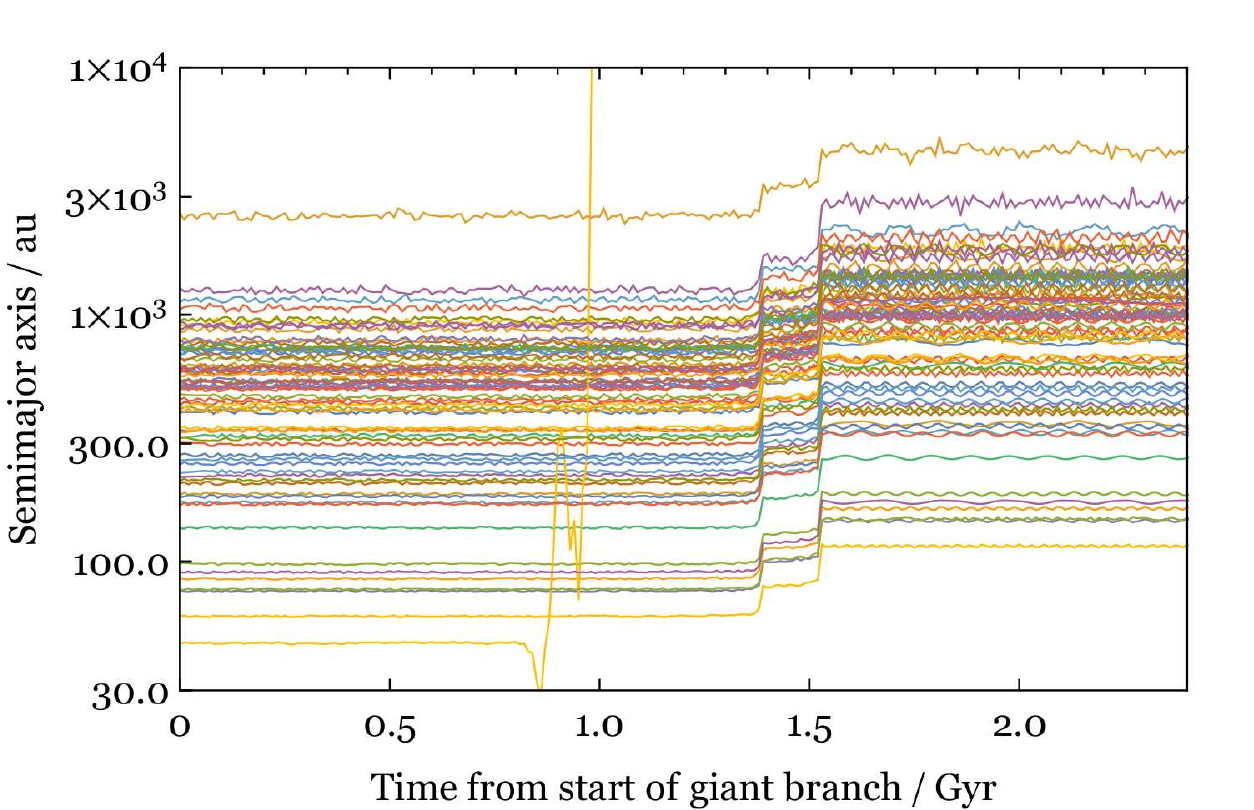}
\ \ \ \ \ \
\includegraphics[width=8.5cm]{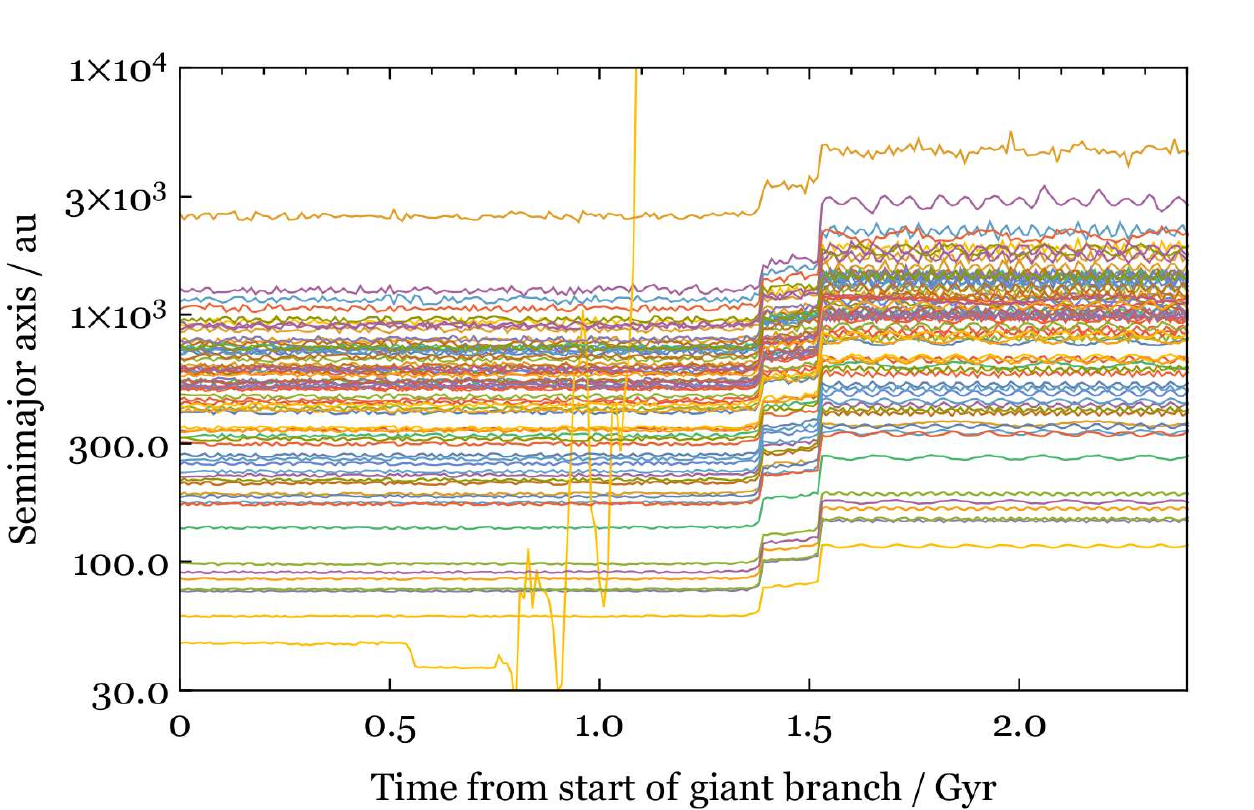}
}
\centerline{}
\centerline{
\includegraphics[width=8.5cm]{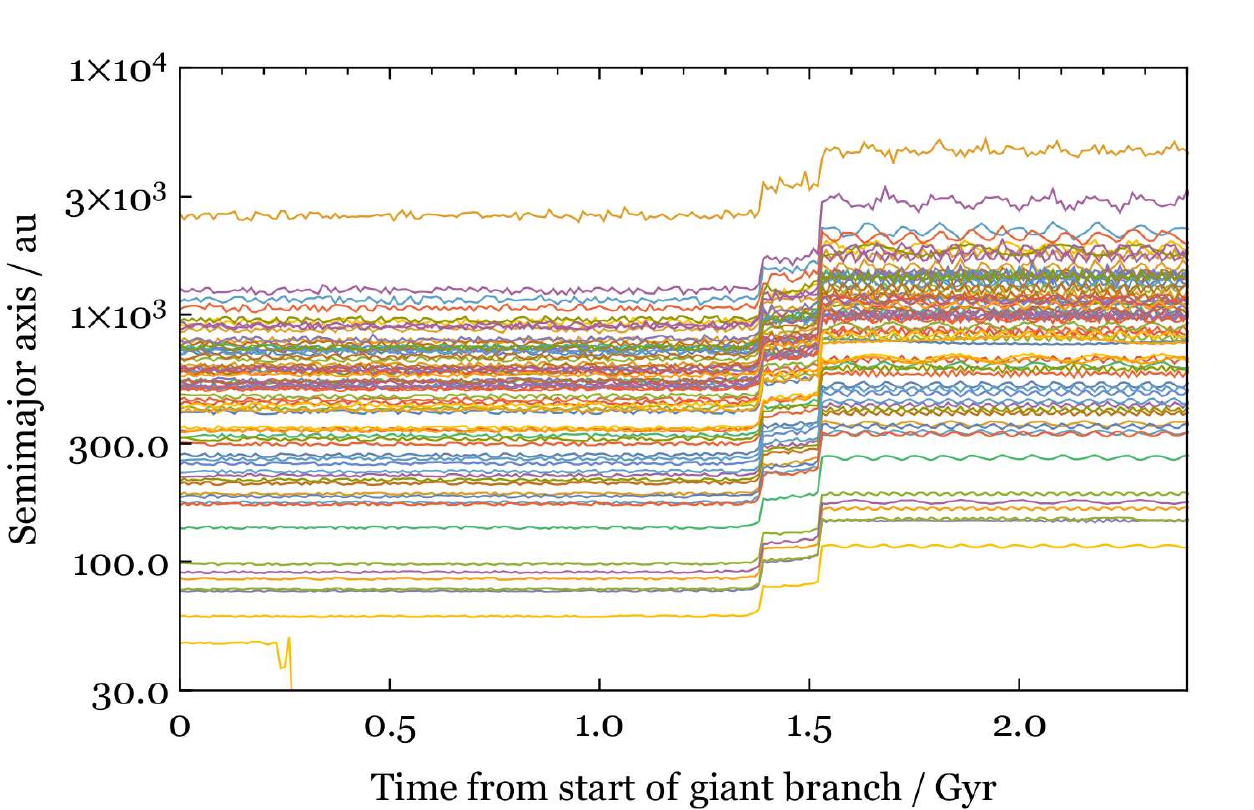}
\ \ \ \ \ \
\includegraphics[width=8.5cm]{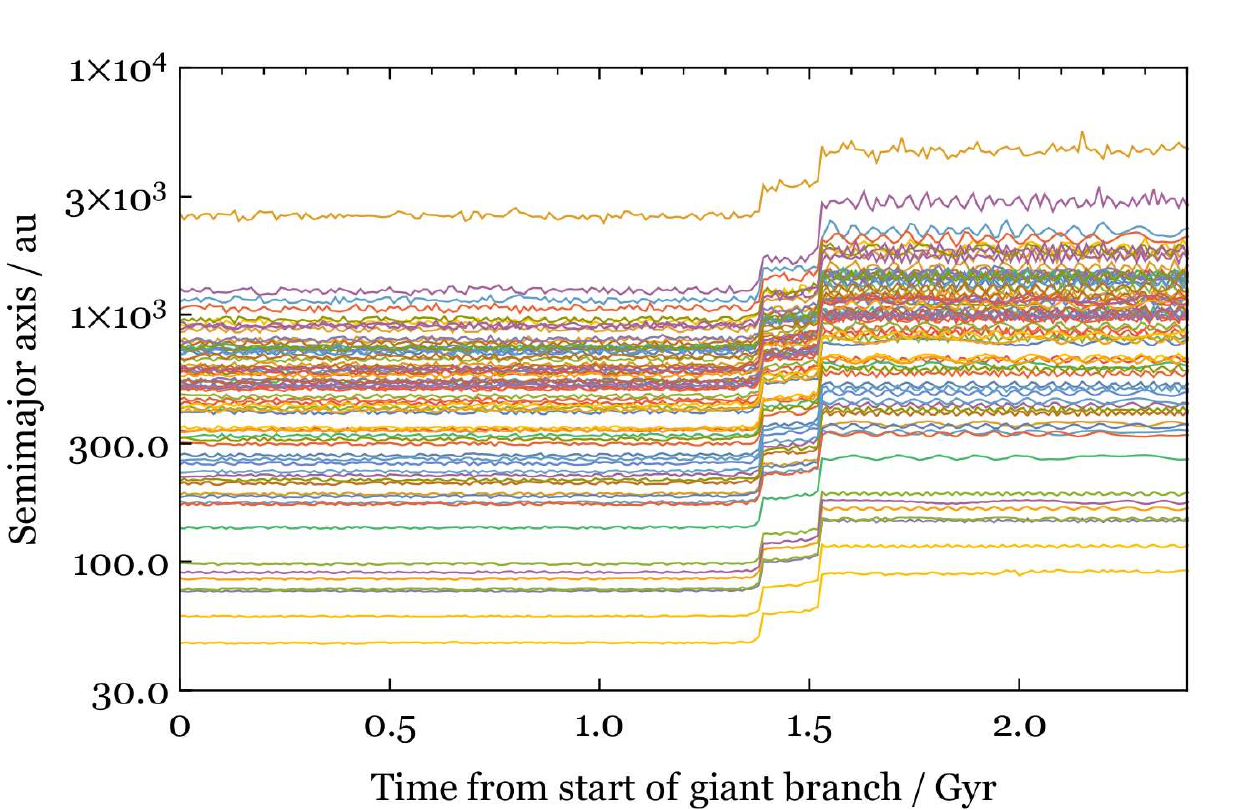}
}
\caption{
Repeat post-main-sequence simulations of the same
70 planetesimals in order to indicate how the chaotic nature
of these systems alters their unstable fraction. In three
of the four systems, one planetesimal becomes unstable,
whereas in the other, all planetesimals remain stable.
}
\label{postMS2}
\end{figure*}
%%%%%%%%%%%%%%%% Figure (dataset 110)

Figure \ref{postMS1} illustrates the results for System \#1.
The doubling of the semimajor axes occurs in two distinct steps,
corresponding to the tips of the red giant and asymptotic giant branch phases
around 1.5 Gyr into the simulation. During this time, an planetesimal's eccentricity 
variation is shifted upwards
by an amount which is correlated to semimajor axis: roughly, for $10^2, 10^3$ and $10^4$ au, 
eccentricity variation is changed respectively to $10^{-3}, 10^{-2}$ and $10^{-1}$.
The inclination remains unaltered by mass loss, an expected result when the mass
loss is isotropic \citep{veretal2013a,doskal2016a,doskal2016b}. Further, the mutual inclination
between the major planets and the planetesimals during the mass loss episode is not expected
to contribute to possible instability \citep{veretal2018}.

The plots in Figure \ref{postMS1} do illustrate some instability, which is to be expected. 
However, only 5 of the 200 planetesimals simulated become unstable, meaning that the
planetesimal disc remains relatively undisturbed, just with double the semimajor axis
and a usually negligible eccentricity shift. 

The occurrence of instability is stochastic. In order to test how the extent of instability
changes due to chaos, in Fig. \ref{postMS2} we show the evolution of a different system
(denoted system \#4) of 70 planetesimals, where the integration was repeated three times with the same initial conditions. However, in each case the integration was interrupted and restarted at different times {\rev during the giant branch evolution}. Each plot exhibits different variations in the oscillations of each curve, and in three of
the four plots, one planetesimal becomes unstable (in the other there is no instability). The
robustness of the evolution in all cases suggests that the variations in instabilities
introduced by stochasticity do not affect our conclusions.

\section{White dwarf evolution}
\label{sec:WD}

Evolution during the white dwarf phase proceeds similarly to that along the main sequence phase
(Section 3), but with a few differences. The largest objects ($\gtrsim 100$ km) in the planetesimal disc now 
occupy a (primarily) $80-2000$~au annulus, from originally 
a $40-1000$~au annulus. Further, the timescale for white dwarf cooling can reach a Hubble time ($>$ 13 Gyr),
although the oldest known metal-polluted white dwarf has a cooling age of about 
8 Gyr \citep{holetal2017,holetal2018}
\footnote{{\rev A white dwarf with a cooling age of 8 Gyr corresponds to a main-sequence progenitor mass larger than the $1.0M_{\odot}$ that we considered here}. Older polluted systems may exist but are difficult to detect because of their faintness.}.

White dwarf luminosities monotonically decrease from the moment they are 
born \citep[e.g.][]{mestel1952}. This fact, coupled with the extended semimajor axes of the planetesimals,
illustrates that the Yarkovsky and YORP effects are negligible for the $\gtrsim 100$ km objects.
We note importantly, however, that boulders and pebbles may still be significantly affected by
this radiation around young white dwarfs, where even the weaker force generated by
Poynting–Robertson drag may crowd the region around the white dwarf with 
debris \citep{stoetal2015,veretal2015c}.

We can estimate the consequences of doubling of the semimajor axis of our planetesimal discs 
with respect to external forces. For Galactic tides, \cite{vereva2013b} defined two different
regimes: ``adiabatic'' and ``nonadiabatic'' (similar to the terms used for mass loss), where 
the adiabatic regime is defined by
when the orbital timescale is faster than the Galactic tidal timescale. They showed how
the equations of motion differ in these two regimes. The planetesimal discs here are
comfortably within the adiabatic regime \citep{veretal2014c}, where Galactic tides
do not alter the semimajor axis $a$, but vary the eccentricity and inclination by an amount proportional
to $a^{3/2}$. Further, for stellar flybys, equation 46 of \cite{vermoe2012} indicates that
the number of flybys which trigger an orbital eccentricity exceeding a given value is, roughly, also
proportional to $a^{3/2}$. Hence, doubling the semimajor axis increases the eccentricity excitation
from external forces approximately by a factor of 2.8. Consequently, for this eccentricity variation
to reach 0.1 during the white dwarf phase, it would need to be at the 0.03 level during the main-sequence.

Such eccentricity variations are important only if they are already on a highly-eccentric orbit or could trigger instability with a sufficiently close
major planet. Instability with the exo-Neptune is the primary driver of orbital architecture variation of 
an exo-Kuiper belt during the white dwarf phase. \cite{bonetal2011} quantified how a single 
planet orbiting a white dwarf could drive
external exo-Kuiper belt objects out of their annulus inward,
and \cite{musetal2018} did the same, but for systems of three major planets.
\cite{musetal2018} demonstrated that in order for these objects to actually reach the
Roche radius of the white dwarf (and hence pollute it), a gravitational instability amongst
multiple major planets is necessary. Instability amongst terrestrial planets provides
a much better match to the observed pollution rate as a function of time \citep{holetal2018} 
than do giant planets (such as the solar system giant planets).

Because Jupiter, Saturn, Uranus and Neptune are expected to remain stable through at least
10 Gyr of Solar white dwarf cooling \citep{lasgas2009,hayetal2010,zeebe2015,veras2016b},
we hence do not expect scattered disc objects such as Sedna to pollute the Solar white dwarf.
The orbital elements of these scattered disc objects will have not significantly changed since the last
scattering event (either amongst planets or from stellar flybys). In solar system analogues 
with no such scattering events, the orbital
elements should reflect the values attained at the end of stellar cluster dispersion.

\section{Discussion}

Our study aimed to achieve a qualitative understanding of how the largest ($\gtrsim 100$ km) 
planetesimal disc objects evolve through time. {\rev We focussed on the initial semimajor axis range $40-1000$~au. Planetesimals likely formed in a more restricted range; as discussed earlier, observations of debris discs indicates that their maximal radial extent is approximately 150 au. Nevertheless, because we adopted test particles, we were able to explore locations exterior to 150 au to which a planetesimal might have been perturbed.}

{\rev Our} results do not profess to answer, but instead provide a foundation for, tackling 
major topical questions, such as: what
is the frequency of interstellar interlopers within the solar system from post-main-sequence ejection
\citep{doetal2018,rafikov2018,moromartin2019}? {\rev (Section 6.1)}, {\rev and} what is the dominant delivery mechanism in metal-polluted white dwarf 
systems \citep{veras2016a}? ({\rev Section 6.2})

Obtaining answers to both questions {\rev requires a comparative population analysis, which we now perform.}

\subsection{Implications for white dwarf pollution}

{\rev
The mass reservoirs which primarily generate white dwarf pollution are still unknown. One candidate reservoir is the population of large exo-Kuiper belt planetesimals, which has been the focus of this study. 

This reservoir is particularly important to consider because \cite{bonetal2011} demonstrated that a Neptune analogue at a 30 au separation from the star could scatter a sufficient amount of exo-Kuiper belt material extending out {\revv from 30 au} to about 48 au towards the white dwarf to explain metal accretion rates. This conclusion is strongly dependent on the belt structure and the time at which the belt was initialized. {\revv Unlike our study, \cite{bonetal2011} modelled planetesimals which were much closer to the planet and hence much more conducive to instabilities due to overlapping resonances. They also} performed integrations along main-sequence evolution for just $10^7$ yr (to dynamically settle the belt), followed by integrations during giant branch evolution.

Hence, placing that study into context is important {\revv when comparing the amount of pollution that has been generated in both studies}. {\revv \cite{bonetal2011} did not perform stellar cluster simulations and we did not consider planetesimals closer to (within 40 au) the exo-Neptune. Further, we note that our nil result for pollution should not imply that planetesimals at or slightly beyond 40 au can never pollute white dwarfs. Major planets could exist beyond 30 au (a notable example being the HR 8799 system; \citealt*{maretal2008,maretal2010}) and could generate the same type of instability that was showcased in \cite{bonetal2011} in those regions to generate pollution.}

%Our investigation instead tracks a belt of large ($\gtrsim 100$ km), more spatially extended ($40-1000$~au) planetesimals from formation to death across all stellar phases. In contrast to \cite{bonetal2011}, our simulations produced no pollution at all. The reasons probably include: (i) the origin of the belt structures in our study arose from stellar cluster evolution and (ii) the belt structure in \cite{bonetal2011} extended inwards to 30 au.  
%Our study also contained truer solar system analogues, including all four giant planets (as opposed to just one), but more major planets might actually enhance pollution through scattering 

{\revv Further,} by considering an annulus of planetesimals extending out to $10^3$~au, we also have probed pollution prospects towards the inner exo-Oort-cloud. Separations out to $10^3$~au have previously been largely unexplored, as previous exo-Oort cloud-based studies have modelled minor planets in the separation range $10^3 - 10^4$ au  \citep{caihey2017}, $1 \times 10^3 - 5 \times 10^4$ au \citep{stoetal2015} and $10^4 - 10^5$ au  \citep{veretal2014b}. 

The numerical integrations in all {\revv these studies of planetesimals beyond $10^3$ au} yielded nonzero quantities of white dwarf pollutants, partially triggered by excitation due to giant branch mass loss \citep{verwya2012,veretal2014b,veretal2014c}. These excited planetesimals evolve non-adiabatically in both the mass loss \citep{veretal2011} and 
Galactic tidal \citep{veretal2014c} senses (meaning that their orbital evolution due to
each force cannot be characterised in a closed explicit form), and hence are actually more likely to strike the white dwarf than planetesimals in our initial separation range of $40-1000$~au\footnote{We note that none of these studies have explicitly explored the sensitive dependence of exo-Oort-cloud evolution on the specific Galactic model adopted \citep{portegieszwart2018,toretal2019} nor exchange mechanisms with both planetesimal discs \citep{weilev1997,shaetal2015,shaetal2019} and the interstellar medium \citep{heitre1986}.}.

We conclude that {\revv the range $40-10^3$ au} represents a ``sweet spot" where we would not expect pollutants dynamically, at least for solar system analogues. Further, chemically, most pollutants are rocky \citep{zucetal2007,ganetal2012,juryou2014,haretal2018,holetal2018,doyetal2019,swaetal2019b,xuetal2019,bonetal2020} rather than volatile-rich \citep{xuetal2017}, suggesting that the main source of pollutants lies inward of 30 au. Despite this chemical suggestion, which has increased in robustness over the last few years, many investigations have dynamically invoked the reservoir of planetesimals initially beyond 30 au as a viable source of white dwarf pollution \citep{bonetal2011,veretal2014b,stoetal2015,caihey2017}, helping to motivate our study.

}

\subsection{Implications for interstellar planetesimals}

{\rev 
Another motivation for our investigation is to explore the effect of ``full-lifetime evolution" on the eventual ejection of planetesimals from an initial $40-1000$~au disc. The flux of such planetesimals in space is unknown, but was predicted to be large enough to produce interstellar interlopers, even before the discovery of 1I/‘Oumuamua \citep{moretal2009}. These planetesimals provide important probes of the outer parts of planetary systems, probes which are not yet available by other means \citep{mcgcha1989,stern1990,wyaetal2017}.

After the discovery of 1I/‘Oumuamua, theoretical investigations of interstellar interlopers blossomed \citep[e.g.][]{doetal2018,jacetal2018,katz2018,2018MNRAS.479L..17P,rafikov2018,rayetal2018a,rayetal2018b,banetal2019,moromartin2018,moromartin2019,malper2020}, ensuring that our study is timely. Because those other studies focus on individual aspects of generating and delivering these planetesimals to the solar system, a comprehensive framework is needed to consolidate the plethora of ideas posed.

We do not claim to provide such a framework. However, what we do, for the first time, is bring together all phases of stellar evolution in a self-consistent way for the $40-1000$ au region. Our findings are similar to our result for white dwarf pollution: dynamical stagnation. Just as our planetesimals do not collide with the star, they also rarely escape after the stellar cluster phase. 

During the cluster phase, we showed that for the $40-1000$~au semimajor axis region, the escape fraction could be any value depending on the physical and orbital properties of the stars in the cluster (in Table \ref{tab:loss}, the escape fraction during the cluster phase ranged from 2.9 per cent to 100 per cent). This result was not {\it a priori} trivial because the planetesimals were initially dynamically cold (on circular and coplanar orbits), and with a maximum semimajor axis that was set at just 1000 au, which is two orders of magnitude smaller than the typical Hill ellipsoid axes of a planetary system \citep{veretal2014c}.

In subsequent phases of stellar evolution, the ejection fraction is comparatively negligible. As we illustrated in Figs. \ref{MS1}-\ref{MS2}, ejection does not occur along the main sequence except during a chance, particularly close flyby. During post-main-sequence evolution, ejection is more likely, but only at the few per cent level (see Fig. \ref{postMS1}), and sometimes not at all (Fig. \ref{postMS2}).

Consequently, we conclude that for the $40-1000$~au range, ejection predominantly occurs during the early cluster phase. Although post-main-sequence origins of interlopers like 1I/‘Oumuamua and 2I/Borisov are possible, in this context tidal disruption events around white dwarfs \citep{rafikov2018,malper2020} may be more likely to occur than ejection during the giant branch phases. 

Despite how our conclusion about ejection occurring primarily during the cluster phase is solely based on $1.0M_{\odot}$ stars with four giant planets, we speculate that this conclusion is robust throughout the Galaxy: although the highest mass white dwarf progenitors would yield the greatest giant branch excitation and ejection potential \citep{veretal2011,veretal2020b}, most stars in the Milky Way are actually less massive than the Sun.
}

\section{Conclusion}

Both metal-polluted white dwarfs and the interstellar planetesimals 1I/‘Oumuamua and 2I/Borisov
provide strong motivation for investigating the full life cycle of minor planets. Here, we
take one step towards achieving this understanding by focusing on large ($\gtrsim 100$ km) planetesimals
in an initial semimajor axis range of $40-1000$~au, which corresponds to a planetesimal disc
or exo-Kuiper belt. 

For the first time, we attempted to link the formative pathways of these belts from their stellar
birth cluster to their fate, when the star leaves the main sequence and eventually transforms
into a white dwarf. We performed this task through a series of arguments and simulations
which encompass the strongest forces on these planetesimals, from stellar radiation to mutual,
external and major planet perturbations in solar system analogues. We found that the planetesimal orbital 
distributions obtained at the end of stellar cluster evolution can be used to predict the evolved form of these 
distributions during the white dwarf phase unless a major gravitational instability (amongst major planets
or with a passing star) occurred in-between. This prediction just entails inflating the planetesimal semimajor axes
by an amount which is inversely proportional to the stellar mass loss; the $40-1000$~au range represents a 
``sweet spot'' where other forces are ineffectual at producing major changes.

\section*{Acknowledgements}

{\rev We thank the referee for their insightful and detailed comments, which have significantly improved the manuscript.}
DV gratefully acknowledges the support of the STFC via an Ernest Rutherford Fellowship (grant ST/P003850/1). KR and RS acknowledge the support of the DFG priority program SPP 1992 ``Exploring the Diversity of Extrasolar Planets'' (SP 345/20-1). FFD and MBNK were supported by the Research Development Fund (grant RDF-16-01-16) of Xi'an Jiaotong-Liverpool University (XJTLU), and MBNK acknowledges support from the National Natural Science Foundation of China (grant 11573004). MXC thanks Santiago Torres and Diptajyoti Mukherjee for insightful discussions. AJM acknowledges funding from the Knut and Alice Wallenberg Foundation (project grant 2014.0017) and the Swedish Research Council (starting grant 2017-04945). AS is supported by funding from the European Research Council under the European Community’s H2020 (2014-2020/ERC Grant Agreement No. 669416 ‘LUCKY STAR’). RS has been supported by National Astronomical Observatories of Chinese Academy of Sciences, Silk Road Project, and by National Natural Science Foundation of China under grant No. 11673032. Star cluster simulations have been done on the GPU accelerated cluster ``kepler'', funded by Volkswagen Foundation grants 84678/84680.

\label{lastpage}

\begin{thebibliography}{999}

\bibitem[Aarseth (1999)]{aarseth1999} Aarseth S. J., 1999, Celest. Mech. Dyn. Astron., 73, 127

\bibitem[Aarseth (2010)]{aarseth2010} Aarseth S. J., 2010, Gravitational N-Body Simulations. Cambridge Monographs on Mathematical Physics. Cambridge University Press, Cambridge.

\bibitem[Adams(2010)]{adams2010} Adams, F.~C.\ 2010, ARA\&A, 48, 47

\bibitem[Adams \& Bloch(2013)]{adablo2013} Adams, F.~C., \& Bloch, A.~M.\ 2013, ApJL, 777, L30
  
\bibitem[Adams et al.(2004)]{adaetal2004} Adams, F.~C., Hollenbach, D., Laughlin, G., et al.\ 2004, ApJ, 611, 360

\bibitem[Alcock et al.(1986)]{alcetal1986} Alcock, C., Fristrom, C.~C., \& Siegelman, R.\ 1986, ApJ, 302, 462 

\bibitem[Anderson et al.(2013)]{andetal2013} Anderson, K.~R., Adams, F.~C., \& Calvet, N.\ 2013, ApJ, 774, 9

\bibitem[Andrews(2020)]{andrews2020} Andrews, S.~M.\ 2020, In Press ARA\&A , arXiv:2001.05007

\bibitem[Antoniadou \& Veras(2016)]{antver2016} Antoniadou, K.~I., \& Veras, D.\ 2016, MNRAS, 463, 4108 

\bibitem[Antoniadou \& Veras(2019)]{antver2019} Antoniadou, K.~I., \& Veras, D.\ 2019, A\&A, 629, A126

\bibitem[Batygin et al.(2020)]{batetal2020} Batygin, K., Adams, F.~C., Batygin, Y.~K., et al.\ 2020, AJ, 159, 101

\bibitem[Belczynski, Kalogera \& Bulik(2002)]{beletal2002} Belczynski K., Kalogera V., Bulik T., 2002, ApJ, 572, 407 

\bibitem[Binney \& Tremaine(2008)]{bintre2008} Binney, J., \& Tremaine, S.\ 2008, Galactic Dynamics: Second Edition

\bibitem[Bland-Hawthorn \& Gerhard(2016)]{BlandHawthornGerhard16} Bland-Hawthorn, J., \& Gerhard, O.\ 2016, ARAA, 54, 529

\bibitem[Bonsor \& Wyatt(2010)]{bonwya2010} Bonsor, A., \& Wyatt, M.\ 2010, MNRAS, 409, 1631

\bibitem[Bonsor et al.(2011)]{bonetal2011} Bonsor, A., Mustill, A.~J., \& Wyatt, M.~C.\ 2011, MNRAS, 414, 930 

\bibitem[Bonsor et al.(2013)]{bonetal2013} Bonsor, A., Kennedy, G.~M., Crepp, J.~R., et al.\ 2013, MNRAS, 431, 3025

\bibitem[Bonsor et al.(2014)]{bonetal2014} Bonsor, A., Kennedy, G.~M., Wyatt, M.~C., et al.\ 2014, MNRAS, 437, 3288

\bibitem[Bonsor \& Veras(2015)]{bonver2015} Bonsor, A., \& Veras, D.\ 2015, MNRAS, 454, 53 

\bibitem[Bonsor et al.(2020)]{bonetal2020} Bonsor, A., Carter, P.~J., Hollands, M., et al.\ 2020, MNRAS In Press, arXiv:2001.04499

\bibitem[Booth et al.(2009)]{boothetal2009} Booth, M., Wyatt, M.~C., Morbidelli, A., et al.\ 2009, MNRAS, 399, 385

\bibitem[Bottke et al.(2005)]{botetal2005} Bottke, W.~F., Durda, D.~D., Nesvorn{\'y}, D., et al.\ 2005, Icarus, 179, 63

\bibitem[Brasser et al.(2006)]{brasseretal2006} Brasser, R., Duncan, M.~J., \& Levison, H.~F.\ 2006, Icarus, 184, 59

\bibitem[Brasser \& Morbidelli(2013)]{bramor2013} Brasser, R., \& Morbidelli, A.\ 2013, Icarus, 225, 40

\bibitem[Brown et al.(2017)]{broetal2017} Brown, J.~C., Veras, D., \& G{\"a}nsicke, B.~T.\ 2017, MNRAS, 468, 1575 

\bibitem[Brucalassi et al.(2017)]{bruetal2017} Brucalassi, A., Koppenhoefer, J., Saglia, R., et al.\ 2017, A\&A, 603, A85

\bibitem[Brunini \& Zanardi(2016)]{bruzan2016} Brunini, A., \& Zanardi, M.\ 2016, MNRAS, 455, 4487

\bibitem[Cai et al.(2015)]{caietal2015} Cai M. X., Meiron Y., Kouwenhoven M. B. N., Assmann P., Spurzem R., 2015, ApJS, 219, 31

\bibitem[Cai et al.(2016)]{caietal2016} Cai M.~X., Gieles M., Heggie D.~C., Varri A.~L., 2016, MNRAS, 455, 596

\bibitem[Cai et al.(2017)]{caietal2017} Cai, M.~X., Kouwenhoven, M.~B.~N., Portegies Zwart, S.~F., et al. 2017, MNRAS, 470, 4337

\bibitem[Cai et al.(2018)]{caietal2018} Cai M. X., Portegies Zwart S., van Elteren A., 2018, MNRAS, 474, 5114

\bibitem[Cai et al.(2019)]{caietal2019} Cai, M.~X., Portegies Zwart, S., Kouwenhoven, M.~B.~N., et al. 2019, MNRAS, 489, 4311

\bibitem[Caiazzo \& Heyl(2017)]{caihey2017} Caiazzo, I., \& Heyl, J.~S.\ 2017, MNRAS, 469, 2750

\bibitem[Carrera et al.(2017)]{Carrera+17} Carrera, D., Gorti, U., Johansen, A., et al.\ 2017, ApJ, 839, 16

\bibitem[Cassan et al.(2012)]{casetal2012} Cassan, A., Kubas, D., Beaulieu, J.-P., et al.\ 2012, Nature, 481, 167 

\bibitem[Chambers(1999)]{chambers1999} Chambers, J.~E.\ 1999, MNRAS, 304, 793 

\bibitem[Chen et al.(2016)]{cheetal2016} Chen, Y.-Y., Ma, Y., \& Zheng, J.\ 2016, MNRAS, 458, 4277

\bibitem[Cibulkov{\'a} et al.(2014)]{cibetal2014} Cibulkov{\'a}, H., Bro{\v{z}}, M., \& Benavidez, P.~G.\ 2014, Icarus, 241, 358

\bibitem[Concha-Ram{\'\i}rez et al.(2019)]{conetal2019} Concha-Ram{\'\i}rez, F., Wilhelm, M.~J.~C., Portegies Zwart, S., et al.\ 2019, MNRAS, 490, 5678

\bibitem[Correa-Otto \& Gil-Hutton(2017)]{corgil2017} Correa-Otto, J.~A., \& Gil-Hutton, R.~A.\ 2017, A\&A, 608, A116

\bibitem[Coutu et al.(2019)]{couetal2019} Coutu, S., Dufour, P., Bergeron, P., et al.\ 2019, ApJ, 885, 74

\bibitem[Dai et al.(2018)]{daietal2018} Dai, Y.-Z., Liu, H.-G., Wu, W.-B., et al.\ 2018, MNRAS, 480, 4080

\bibitem[Debes \& Sigurdsson(2002)]{debsig2002} Debes, J.~H., \& Sigurdsson, S.\ 2002, ApJ, 572, 556 

\bibitem[Debes et al.(2012)]{debetal2012} Debes, J.~H., Walsh, K.~J., \& Stark, C.\ 2012, ApJ, 747, 148 

\bibitem[Dell'Oro et al.(2012)]{deletal2012} Dell'Oro, A., Cellino, A., \& Paolicchi, P.\ 2012, MNRAS, 425, 1492

\bibitem[Dennihy et al.(2018)]{denetal2018} Dennihy, E., Clemens, J.~C., Dunlap, B.~H., Fanale, S.~M., Fuchs, J.~T., Hermes, J.~J.\ 2018, ApJ, 854, 40 

\bibitem[Do et al.(2018)]{doetal2018} Do, A., Tucker, M.~A., \& Tonry, J.\ 2018, ApJL, 855, L10

\bibitem[Dohnanyi(1969)]{dohnanyi1969} Dohnanyi, J.~S.\ 1969, JGR, 74, 2531

\bibitem[Dong et al.(2010)]{donetal2010} Dong, R., Wang, Y., Lin, D.~N.~C., et al.\ 2010, ApJ, 715, 1036

\bibitem[Dosopoulou \& Kalogera(2016a)]{doskal2016a} Dosopoulou, F., \& Kalogera, V.\ 2016a, ApJ, 825, 70 

\bibitem[Dosopoulou \& Kalogera(2016b)]{doskal2016b} Dosopoulou, F., \& Kalogera, V.\ 2016b, ApJ, 825, 71 

\bibitem[Doyle et al.(2019)]{doyetal2019} Doyle, A.~E., Young, E.~D., Klein, B., et al.\ 2019, Science, 366, 356

\bibitem[Duvvuri et al.(2019)]{duvetal2019} Duvvuri, G., Redfield, S. \& Veras, D. 2019, Submitted to AAS Journals

\bibitem[Ernst, Just \& Spurzem(2009)]{ernetal2009} Ernst A., Just A., Spurzem R., 2009, MNRAS, 399, 141


\bibitem[Farihi(2016)]{farihi2016} Farihi, J.\ 2016, New Astronomy Reviews, 71, 9

\bibitem[Feltzing et al.(2019)]{Feltzing+19} Feltzing, S., Bowers, J.~B., \& Agertz, O.\ 2019, arXiv e-prints, arXiv:1907.08011

\bibitem[Ferlet et al.(1987)]{ferletetal1987} Ferlet R., Hobbs L.~M., Madjar A.~V., 1987, A\&A, 185, 267

\bibitem[Flammini Dotti et al.(2019)]{flaetal2019} Flammini Dotti, F., Kouwenhoven, M.~B.~N., Cai, M.~X., et al.\ 2019, MNRAS, 489, 2280

\bibitem[Flammini Dotti et al.(2020)]{flaetal2020} Flammini Dotti F., Cai M. X., Spurzem R., Kouwenhoven M. B. N., 2020, Origins: From the Protosun to the First Steps of Life. Proceedings of the International Astronomical Union, Volume 345, pp. 293-294

\bibitem[Frankel et al.(2018)]{Frankel+18} Frankel, N., Rix, H.-W., Ting, Y.-S., et al.\ 2018, ApJ, 865, 96

\bibitem[Frewen \& Hansen(2014)]{frehan2014} Frewen, S.~F.~N., \& Hansen, B.~M.~S.\ 2014, MNRAS, 439, 2442 

\bibitem[Gallet et al.(2017)]{galetal2017} Gallet, F., Bolmont, E., Mathis, S., et al.\ 2017, A\&A, 604, A112

\bibitem[G{\"a}nsicke et al.(2006)]{gaeetal2006} G{\"a}nsicke, B.~T., Marsh, T.~R., Southworth, J., \& Rebassa-Mansergas, A.\ 2006, Science, 314, 1908 

\bibitem[G{\"a}nsicke et al.(2012)]{ganetal2012} G{\"a}nsicke, B.~T., Koester, D., Farihi, J., et al.\ 2012, MNRAS, 424, 333

\bibitem[Garc{\'\i}a-S{\'a}nchez et al.(2001)]{garetal2001} Garc{\'\i}a-S{\'a}nchez, J., Weissman, P.~R., Preston, R.~A., et al.\ 2001, A\&A, 379, 634

\bibitem[G{\'a}sp{\'a}r \& Rieke(2014)]{gasrie2014} G{\'a}sp{\'a}r, A., \& Rieke, G.~H.\ 2014, ApJ, 784, 33

\bibitem[Geller et al.(2008)]{geletal2008} Geller, A.~M., Mathieu, R.~D., Harris, H.~C., et al.\ 2008, AJ, 135, 2264

\bibitem[Gomes et al.(2008)]{gometal2008} Gomes, R.~S., Fern{\'a}ndez, J.~A., Gallardo, T., et al.\ 2008, The Solar System Beyond Neptune, 259

\bibitem[Graham et al.(1990)]{graetal1990} Graham, J.~R., Matthews, K., Neugebauer, G., \& Soifer, B.~T.\ 1990, ApJ, 357, 216 

\bibitem[Grishin \& Veras(2019)]{griver2019} Grishin, E., \& Veras, D.\ 2019, MNRAS, 489, 168

\bibitem[Gurri et al.(2017)]{guretal2017} Gurri, P., Veras, D., \& G{\"a}nsicke, B.~T.\ 2017, MNRAS, 464, 321 

\bibitem[Guzik et al.(2019)]{guzetal2019} Guzik, P., Drahus, M., Rusek, K., et al.\ 2019, Nature Astronomy, 467

\bibitem[Hadjidemetriou(1963)]{hadjidemetriou1963} Hadjidemetriou, J.~D.\ 1963, Icarus, 2, 440 

\bibitem[\protect\citeauthoryear{Hamers \& Portegies Zwart}{2016}]{hampor2016} Hamers A.~S., Portegies Zwart S.~F., 2016, MNRAS, 462, L84

\bibitem[Hamers \& Tremaine(2017)]{hamtre2017} Hamers, A.~S., \& Tremaine, S.\ 2017, AJ, 154, 272

\bibitem[Hands et al.(2019)]{hanetal2019} Hands, T.~O., Dehnen, W., Gration, A., et al.\ 2019, MNRAS, 490, 21

\bibitem[Hao et al.(2013)]{haoetal2013} Hao W., Kouwenhoven M.~B.~N., Spurzem R., 2013, MNRAS, 433, 867

\bibitem[Harrison et al.(2018)]{haretal2018} Harrison, J.~H.~D., Bonsor, A., \& Madhusudhan, N.\ 2018, MNRAS, 479, 3814

\bibitem[Hayes et al.(2010)]{hayetal2010} Hayes, W.~B., Malykh, A.~V., \& Danforth, C.~M.\ 2010, MNRAS, 407, 1859

\bibitem[Heisler \& Tremaine(1986)]{heitre1986} Heisler, J., \& Tremaine, S.\ 1986, Icarus, 65, 13 

\bibitem[Hobbs et al.(2005)]{hobetal2005} Hobbs G., Lorimer D.~R., Lyne A.~G., Kramer M., 2005, MNRAS, 360, 974

\bibitem[Hollands et al.(2017)]{holetal2017} Hollands, M.~A., Koester, D., Alekseev, V., Herbert, E.~L., \& G{\"a}nsicke, B.~T.\ 2017, MNRAS, 467, 4970 

\bibitem[Hollands et al.(2018)]{holetal2018} Hollands, M.~A., G{\"a}nsicke, B.~T., \& Koester, D.\ 2018, MNRAS, 477, 93.

\bibitem[Holsapple(2007)]{hols2007} Holsapple, K.~A. 2007, Icarus, 187, 500 

\bibitem[Hurley et al.(2000)]{huretal2000} Hurley, J.~R., Pols, O.~R., \& Tout, C.~A.\ 2000, MNRAS, 315, 543 

\bibitem[Hurley et al.(2001)]{huretal2001} Hurley J.~R., Tout C.~A., Aarseth S.~J., Pols O.~R., 2001, MNRAS, 323, 630

\bibitem[Jackson et al.(2018)]{jacetal2018} Jackson, A.~P., Tamayo, D., Hammond, N., et al.\ 2018, MNRAS, 478, L49

\bibitem[Jura \& Young(2014)]{juryou2014} Jura, M., \& Young, E.~D.\ 2014, Annual Review of Earth and Planetary Sciences, 42, 45

\bibitem[Jutzi et al.(2017)]{jutetal2017} Jutzi, M., Benz, W., Toliou, A., et al.\ 2017, A\&A, 597, A61

\bibitem[Katz(2018)]{katz2018} Katz, J.~I.\ 2018, MNRAS, 478, L95

\bibitem[Kenyon \& Bromley(2008)]{keny2008} Kenyon S.~J., Bromley B.~C., 2008, ApJS, 179, 451

\bibitem[King(1962)]{kin1962} King I., 1962, AJ, 67, 471 

\bibitem[Kobayashi \& L{\"o}hne(2014)]{koba2014} Kobayashi H., L{\"o}hne T., 2014, MNRAS, 442, 3266

\bibitem[Koester et al.(2014)]{koeetal2014} Koester, D., G{\"a}nsicke, B.~T., \& Farihi, J.\ 2014, A\&A, 566, A34 

\bibitem[Kokubo, Yoshinaga \& Makino(1998)]{koketal1998} Kokubo E., Yoshinaga K., Makino J., 1998, MNRAS, 297, 1067

\bibitem[Krivov et al.(2013)]{krietal2013} Krivov, A.~V., Eiroa, C., L{\"o}hne, T., et al.\ 2013, ApJ, 772, 32

\bibitem[Krivov et al.(2018)]{krietal2018} Krivov A.~V., Ide A., L{\"o}hne T., Johansen A., Blum J., 2018, MNRAS, 474, 2564

\bibitem[Kunitomo et al.(2011)]{kunetal2011} Kunitomo, M., Ikoma, M., Sato, B., Katsuta, Y., \& Ida, S.\ 2011, ApJ, 737, 66 

\bibitem[Kroupa(2001)]{kroupa2001} Kroupa P., 2001, MNRAS, 322, 231

\bibitem[Laskar \& Gastineau(2009)]{lasgas2009} Laskar, J., \& Gastineau, M.\ 2009, Nature, 459, 817

\bibitem[Lawler et al. (2017)]{lawleretal2017} Lawler S.~M., Shankman C., Kaib N., Bannister M.~T., Gladman B., Kavelaars J.~J., 2017, AJ, 153, 33

\bibitem[Le{\~a}o et al.(2018)]{leaetal2018} Le{\~a}o, I.~C., Canto Martins, B.~L., Alves, S., et al.\ 2018, A\&A, 620, A139

\bibitem[Li et al.(2019)]{lietal2019} Li, D., Mustill, A.~J., \& Davies, M.~B.\ 2019, MNRAS, 488, 1366

\bibitem[Liou \& Kaufmann(2008)]{liokau2008} Liou, J.-C., \& Kaufmann, D.~E.\ 2008, The Solar System Beyond Neptune, 425

\bibitem[L{\"o}hne et al.(2008)]{lohetal2008} L{\"o}hne, T., Krivov, A.~V., \& Rodmann, J.\ 2008, ApJ, 673, 1123

\bibitem[Madappatt et al.(2016)]{madetal2016} Madappatt, N., De Marco, O., \& Villaver, E.\ 2016, MNRAS, 463, 1040 

\bibitem[Makarov \& Veras(2019)]{makver2019} Makarov, V.~V., \& Veras, D.\ 2019, ApJ, 886, 127

\bibitem[Malamud \& Perets(2020)]{malper2020} Malamud, U., \& Perets, H.~B.\ 2020, MNRAS, 129

\bibitem[Malhotra(1993)]{Malhotra1993} Malhotra, R.\ 1993, Nature, 365, 819

\bibitem[Malmberg et al.(2007)]{maletal2007} Malmberg, D., de Angeli, F., Davies, M.~B., et al.\ 2007, MNRAS, 378, 1207

\bibitem[Malmberg et al.(2011)]{maletal2011} Malmberg, D., Davies, M.~B., \& Heggie, D.~C.\ 2011, MNRAS, 411, 859

\bibitem[Mann et al.(2017)]{manetal2017} Mann, A.~W., Gaidos, E., Vanderburg, A., et al.\ 2017, AJ, 153, 64

\bibitem[Manser et al.(2019)]{manetal2019} Manser, C.~J., G{\"a}nsicke, B.~T., Eggl, S., et al.\ 2019, Science, 364, 66

\bibitem[Marois et al.(2008)]{maretal2008} Marois, C., Macintosh, B., Barman, T., et al.\ 2008, Science, 322, 1348

\bibitem[Marois et al.(2010)]{maretal2010} Marois, C., Zuckerman, B., Konopacky, Q.~M., et al.\ 2010, Nature, 468, 1080

\bibitem[Martin et al.(2020)]{maretal2020} Martin, R.~G., Livio, M., Smallwood, J.~L., et al.\ 2020, MNRAS, In Press

\bibitem[McGlynn \& Chapman(1989)]{mcgcha1989} McGlynn, T.~A., \& Chapman, R.~D.\ 1989, ApJL, 346, L105

\bibitem[McMillan et al.(2012)]{mcmillan2012} McMillan S., Portegies Zwart S., van Elteren A., Whitehead A., 2012, in Capuzzo-Dolcetta R., Limongi M., Tornambe` A., eds, ASP Conf. Ser. Vol. 453, Advances in Computational Astrophysics: Methods, Tools, and Outcome. Astron. Soc. Pac., San Francisco, p. 129

\bibitem[Meech et al.(2017)]{meeetal2017} Meech, K.~J., Weryk, R., Micheli, M., et al.\ 2017, Nature, 552, 378

\bibitem[Mestel(1952)]{mestel1952} Mestel, L.\ 1952, MNRAS, 112, 583

\bibitem[Miller(1964)]{mil1964} Miller R.~H., 1964, ApJ, 140, 250

\bibitem[Milone et al.(2012)]{miletal2012} Milone, A.~P., Piotto, G., Bedin, L.~R., et al.\ 2012, A\&A, 540, A16

\bibitem[Minchev et al.(2013)]{Minchev+13} Minchev, I., Chiappini, C., \& Martig, M.\ 2013, A\&A, 558, A9

\bibitem[Minchev et al.(2018)]{Minchev+18} Minchev, I., Anders, F., Recio-Blanco, A., et al.\ 2018, MNRAS, 481, 1645

\bibitem[Moore et al.(2016)]{mooetal2016} Moore, J.~M., McKinnon, W.~B., Spencer, J.~R., et al.\ 2016, Science, 351, 1284

\bibitem[Morbidelli et al.(2007)]{moretal2007} Morbidelli, A., Tsiganis, K., Crida, A., et al.\ 2007, AJ, 134, 1790

\bibitem[Morbidelli et al.(2018)]{morbidelli2018} Morbidelli, A., Nesvorny, D., Laurenz, V., et al.\ 2018, Icarus, 305, 262

\bibitem[Morbidelli \& Nesvorny(2019)]{mornes2019} Morbidelli, A., \& Nesvorny, D.\ 2019, arXiv:1904.02980, Review chapter to be published in the book ``The Transneptunian Solar System'', Eds: Dina Prialnik, Maria Antonietta Barucci, Leslie Young, Elsevier 

\bibitem[Moro-Mart{\'\i}n et al.(2009)]{moretal2009} Moro-Mart{\'\i}n, A., Turner, E.~L., \& Loeb, A.\ 2009, ApJ, 704, 733

\bibitem[Moro-Mart{\'\i}n(2018)]{moromartin2018} Moro-Mart{\'\i}n, A.\ 2018, ApJ, 866, 131

\bibitem[Moro-Mart{\'\i}n(2019)]{moromartin2019} Moro-Mart{\'\i}n, A.\ 2019, AJ, 157, 86

\bibitem[Mustill \& Villaver(2012)]{musvil2012} Mustill, A.~J., \& Villaver, E.\ 2012, ApJ, 761, 121 

\bibitem[Mustill et al.(2013)]{musetal2013} Mustill, A.~J., Marshall, J.~P., Villaver, E., et al.\ 2013, MNRAS, 436, 2515 

\bibitem[Mustill et al.(2014)]{musetal2014} Mustill, A.~J., Veras, D., \& Villaver, E.\ 2014, MNRAS, 437, 1404 

\bibitem[Mustill et al.(2018)]{musetal2018} Mustill, A.~J., Villaver, E., Veras, D.,  G{\"a}nsicke, B.~T., Bonsor, A. \ 2018, MNRAS, 476, 3939.

\bibitem[Nesvorn{\'y} et al.(2011)]{nesetal2011} Nesvorn{\'y}, D., Vokrouhlick{\'y}, D., Bottke, W.~F., et al.\ 2011, AJ, 141, 159

\bibitem[Nesvorn{\'y} \& Morbidelli(2012)]{nesmor2012} Nesvorn{\'y}, D., \& Morbidelli, A.\ 2012, AJ, 144, 117

\bibitem[Nesvorn{\'y} \& Vokrouhlick{\'y}(2016)]{nesvok2016} Nesvorn{\'y}, D., \& Vokrouhlick{\'y}, D.\ 2016, ApJ, 825, 94

\bibitem[Nesvorn{\'y}(2018)]{nesvorny2018} Nesvorn{\'y}, D.\ 2018, ARA\&A, 56, 137

\bibitem[Nesvorn{\'y} et al.(2019)]{nesetal2019} Nesvorn{\'y}, D., Li, R., Youdin, A.~N., et al.\ 2019, Nature Astronomy, 3, 808

\bibitem[Nicholson et al.(2019)]{nicetal2019} Nicholson, R.~B., Parker, R.~J., Church, R.~P., et al.\ 2019, MNRAS, 485, 4893

\bibitem[Nordhaus \& Spiegel(2013)]{norspi2013} Nordhaus, J., \& Spiegel, D.~S.\ 2013, MNRAS, 432, 500 

\bibitem[{\"O}berg \& Wordsworth(2019)]{ObergWordsworth2019} {\"O}berg, K.~I., \& Wordsworth, R.\ 2019, AJ, 158, 194

\bibitem[Omarov(1962)]{omarov1962} Omarov, T.~B. 1962, Izv. Astrofiz. Inst. Acad. Nauk. KazSSR, 14, 66

\bibitem['Oumuamua ISSI Team et al.(2019)]{banetal2019} 'Oumuamua ISSI Team, Bannister, M.~T., Bhandare, A., et al.\ 2019, Nature Astronomy, 3, 594

\bibitem[Parker(2015)]{parker2015} Parker, A.~H.\ 2015, Icarus, 247, 112

\bibitem[Parravano et al.(2011)]{paretal2011} Parravano, A., McKee, C.~F., \& Hollenbach, D.~J.\ 2011, ApJ, 726, 27

\bibitem[Pascucci \& Tachibana(2010)]{pascuccitachibana2010} Pascucci, I., \& Tachibana, S.\ 2010, Protoplanetary Dust: Astrophysical and Cosmochemical Perspectives, 263

\bibitem[Pelupessy et al.(2013)]{pelupessy2013} Pelupessy F. I., van Elteren A., de Vries N., McMillan S. L. W., Drost N., Portegies Zwart S. F., 2013, A\&A ,557, A84

\bibitem[Petrovich \& Mu{\~n}oz(2017)]{petmun2017} Petrovich, C., \& Mu{\~n}oz, D.~J.\ 2017, ApJ, 834, 116 

\bibitem[Pfalzner et al.(2018)]{pfaetal2018} Pfalzner, S., Bhandare, A., Vincke, K., et al.\ 2018, ApJ, 863, 45

\bibitem[Pirani et al.(2019)]{Pirani+19} Pirani, S., Johansen, A., Bitsch, B., et al.\ 2019, A\&A, 623, A169

\bibitem[Plummer(1911)]{plummer1911} Plummer H.~C., 1911, MNRAS, 71, 460

\bibitem[Polishook et al.(2017)]{poletal2017} Polishook, D., Moskovitz, N., Thirouin, A., et al.\ 2017, Icarus, 297, 126

\bibitem[Portegies Zwart(2009)]{2009ApJ...696L..13P} Portegies Zwart S.~F., 2009, ApJL, 696, L13

\bibitem[Portegies Zwart et al.(2011)]{portegieszwart2011} AMUSE: Astrophysical Multipurpose Software Environment, Astrophysics Source Code Library, record (ascl:1107.007)

\bibitem[Portegies Zwart(2013)]{portegieszwart2013} Portegies Zwart, S.\ 2013, MNRAS, 429, L45 

\bibitem[Portegies Zwart \& J{\'\i}lkov{\'a}(2015)]{2015MNRAS.451..144P} Portegies Zwart S.~F., J{\'\i}lkov{\'a} L., 2015, MNRAS, 451, 144

\bibitem[Portegies Zwart et al. (2018)]{2018MNRAS.479L..17P} Portegies Zwart S., Torres S., Pelupessy I., B{\'e}dorf J., Cai M.~X., 2018, MNRAS, 479, L17

\bibitem[Portegies Zwart et al.(2018)]{portegieszwart2018} Portegies Zwart, S., Torres, S., Pelupessy, I., et al.\ 2018, MNRAS, 479, L17

\bibitem[Portegies Zwart(2019)]{2019A&A...622A..69P} Portegies Zwart S., 2019, A\&A, 622, A69

\bibitem[Portell de Mora et al. (2011)]{portell2011} Portell de Mora J., Garc{\'\i}a-Berro E., Estepa C., Casta{\~n}eda J., Clotet M., 2011, SPIE,  818305, SPIE.8183

\bibitem[Punzo, Capuzzo-Dolcetta \& Portegies Zwart, 2014]{2014MNRAS.444.2808P} Punzo D., Capuzzo-Dolcetta R., Portegies Zwart S., 2014, MNRAS, 444, 2808

\bibitem[Quinlan \& Tremaine(1992)]{quitre1992} Quinlan G.~D., Tremaine S., 1992, MNRAS, 259, 505

\bibitem[Rafikov(2018)]{rafikov2018} Rafikov, R.~R.\ 2018, ApJ, 861, 35

\bibitem[Raghavan et al.(2010)]{ragetal2010} Raghavan, D., McAlister, H.~A., Henry, T.~J., et al.\ 2010, ApJS, 190, 1

\bibitem[Rein \& Liu (2012)]{rei2012} Rein H., Liu S.-F., 2012, A\&A, 537, A128

\bibitem[\protect\citeauthoryear{Rao et al.}{2018}]{raoetal2018} Rao S., et al., 2018, A\&A, 618, A18

\bibitem[Rappaport et al.(2018)]{rappaportetal2018} Rappaport, S., Vanderburg, A., Jacobs, T., et al.\ 2018, MNRAS, 474, 1453

\bibitem[Raymond et al.(2018a)]{rayetal2018a} Raymond, S.~N., Armitage, P.~J., Veras, D., et al.\ 2018a, MNRAS, 476, 3031

\bibitem[Raymond et al.(2018b)]{rayetal2018b} Raymond, S.~N., Armitage, P.~J., \& Veras, D.\ 2018b, ApJL, 856, L7

\bibitem[Schr{\"o}der \& Smith(2008)]{schcon2008} Schr{\"o}der, K.-P., \& Smith, R.C.\ 2008, MNRAS, 386, 155 

\bibitem[Sellwood \& Binney(2002)]{SellwoodBinney2002} Sellwood, J.~A., \& Binney, J.~J.\ 2002, MNRAS, 336, 785

\bibitem[Shannon \& Wu(2011)]{shawu2011} Shannon, A., \& Wu, Y.\ 2011, ApJ, 739, 36

\bibitem[Shannon et al.(2015)]{shaetal2015} Shannon, A., Jackson, A.~P., Veras, D., et al.\ 2015, MNRAS, 446, 2059

\bibitem[Shannon \& Dawson(2018)]{shadaw2018} Shannon A., \& Dawson R., 2018, MNRAS, 480, 1870

\bibitem[Shannon et al.(2019)]{shaetal2019} Shannon, A., Jackson, A.~P., \& Wyatt, M.~C.\ 2019, MNRAS, 485, 5511

\bibitem[Shara et al.(2016)]{shaetal2016} Shara, M.~M., Hurley, J.~R., \& Mardling, R.~A.\ 2016, ApJ, 816, 59

\bibitem[Sibthorpe et al.(2018)]{sibetal2018} Sibthorpe, B., Kennedy, G.~M., Wyatt, M.~C., et al.\ 2018, MNRAS, 475, 3046

\bibitem[Silsbee \& Tremaine(2018)]{siltre2018} Silsbee, K., \& Tremaine, S.\ 2018, AJ, 155, 75

\bibitem[Smallwood et al.(2018)]{smaetal2018} Smallwood, J.~L., Martin, R.~G., Livio, M., \& Lubow, S.~H.\ 2018, MNRAS, 480, 57

\bibitem[Smallwood et al.(2019)]{smaetal2019} Smallwood, J.~L., Martin, R.~G., Livio, M., Lubow, S.~H., \& Veras, D.\ 2019, Submitted to MNRAS

\bibitem[Spurzem (1999)]{spurzem1999} Spurzem R., 1999, J. Comput. Appl. Math., 109, 407

\bibitem[Spurzem, et al.(2009)]{spuetal2009} Spurzem R., Giersz M., Heggie D.~C., Lin D.~N.~C., 2009, ApJ, 697, 458

\bibitem[Staff et al.(2016)]{staetal2016} Staff, J.~E., De Marco, O., Wood, P., Galaviz, P., \& Passy, J.-C.\ 2016, MNRAS, 458, 832 

\bibitem[Stern(1990)]{stern1990} Stern, S.~A.\ 1990, PASP, 102, 793

\bibitem[Stern et al.(2015)]{steetal2015} Stern, S.~A., Bagenal, F., Ennico, K., et al.\ 2015, Science, 350, aad1815

\bibitem[Stern et al.(2019)]{steetal2019} Stern, S.~A., Weaver, H.~A., Spencer, J.~R., et al.\ 2019, Science, 364, aaw9771

\bibitem[Stone et al.(2015)]{stoetal2015} Stone, N., Metzger, B.~D., \& Loeb, A.\ 2015, MNRAS, 448, 188 

\bibitem[\protect\citeauthoryear{Sun et al.}{2018}]{sunetal2018} Sun M., Arras P., Weinberg N.~N., Troup N.~W., Majewski S.~R., 2018, MNRAS, 481, 4077

\bibitem[Swan et al.(2019a)]{swaetal2019a} Swan, A., Farihi, J., \& Wilson, T.~G.\ 2019a, MNRAS, 484, L109

\bibitem[Swan et al.(2019b)]{swaetal2019b} Swan, A., Farihi, J., Koester, D., et al.\ 2019b, MNRAS, 490, 202

\bibitem[Tapamo (2009)]{tapamo2009} Tapamo H., 2009, in Gracia J., de Colle F., Downes T., eds, Lecture Notes in Physics, Vol. 791, Jets From Young Stars V. Springer Verlag , Berlin, p. 3

\bibitem[Thommes et al.(2002)]{thoetal2002} Thommes, E.~W., Duncan, M.~J., \& Levison, H.~F.\ 2002, AJ, 123, 2862

\bibitem[Tiscareno \& Malhotra(2009)]{tiscarenoetal2009} Tiscareno, M.~S., \& Malhotra, R.\ 2009, AJ, 138, 827

\bibitem[Torres et al.(2019)]{toretal2019} Torres, S., Cai, M.~X., Brown, A.~G.~A., et al.\ 2019, A\&A, 629, A139

\bibitem[Tsiganis et al.(2005)]{tsietal2005} Tsiganis, K., Gomes, R., Morbidelli, A., et al.\ 2005, Nature, 435, 459

\bibitem[van Elteren et al.(2019)]{flybyetal2019} van Elteren, A., Portegies Zwart, S., Pelupessy, I., et al.\ 2019, A\&A, 624, A120

\bibitem[Vanderbosch et al.(2019)]{vanetal2019} Vanderbosch, Z., Hermes, J.~J., Dennihy, E., et al.\ 2019, Submitted to ApJL, arXiv:1908.09839

\bibitem[Vanderburg et al.(2015)]{vanetal2015} Vanderburg, A., Johnson, J.~A., Rappaport, S., et al.\ 2015, Nature, 526, 546 

\bibitem[Veras \& Armitage(2004)]{verarm2004} Veras, D., \& Armitage, P.~J.\ 2004, MNRAS, 347, 613

\bibitem[Veras et al.(2011)]{veretal2011} Veras, D., Wyatt, M.~C., Mustill, A.~J., Bonsor, A., \& Eldridge, J.~J.\ 2011, MNRAS, 417, 2104 

\bibitem[Veras \& Moeckel(2012)]{vermoe2012} Veras, D., \& Moeckel, N.\ 2012, MNRAS, 425, 680 

\bibitem[Veras \& Wyatt(2012)]{verwya2012} Veras, D., \& Wyatt, M.~C.\ 2012, MNRAS, 421, 2969 

\bibitem[Veras et al.(2013a)]{veretal2013a} Veras, D., Hadjidemetriou, J.~D., \& Tout, C.~A.\ 2013a, MNRAS, 435, 2416 

\bibitem[Veras et al.(2013b)]{veretal2013b} Veras, D., Mustill, A.~J., Bonsor, A., \& Wyatt, M.~C.\ 2013b, MNRAS, 431, 1686 

\bibitem[Veras \& Evans(2013a)]{vereva2013a} Veras, D., \& Evans, N.~W.\ 2013a, MNRAS, 430, 403 

\bibitem[Veras \& Evans(2013b)]{vereva2013b} Veras, D., \& Evans, N.~W.\ 2013b, CeMDA, 115, 123 

\bibitem[Veras et al.(2014a)]{veretal2014a} Veras, D., Jacobson, S.~A., G\"{a}nsicke, B.~T.\ 2014a, MNRAS, 445, 2794 

\bibitem[Veras et al.(2014b)]{veretal2014b} Veras, D., Shannon, A., G\"{a}nsicke, B.~T.\ 2014b, MNRAS, 445, 4175 

\bibitem[Veras et al.(2014c)]{veretal2014c} Veras, D., Evans, N.~W., Wyatt, M.~C., \& Tout, C.~A.\ 2014c, MNRAS, 437, 1127

\bibitem[Veras \& G\"{a}nsicke(2015)]{vergae2015} Veras, D., G\"{a}nsicke, B.~T.\ 2015, MNRAS, 447, 1049 

\bibitem[Veras et al.(2015a)]{veretal2015a} Veras, D., Eggl, S., G{\"a}nsicke, B.~T.\ 2015a, MNRAS, 452, 1945

\bibitem[Veras et al.(2015b)]{veretal2015b} Veras, D., Eggl, S., G{\"a}nsicke, B.~T.\ 2015b, MNRAS, 451, 2814 

\bibitem[Veras et al.(2015c)]{veretal2015c} Veras, D., Leinhardt, Z.~M., Eggl, S., G{\"a}nsicke, B.~T.\ 2015c, MNRAS, 451, 3453 

\bibitem[Veras(2016a)]{veras2016a} Veras, D.\ 2016a, Royal Society Open Science, 3, 150571 

\bibitem[Veras(2016b)]{veras2016b} Veras, D.\ 2016b, MNRAS, 463, 2958 

\bibitem[Veras et al.(2016a)]{veretal2016a} Veras, D., Marsh, T.~M., G\"{a}nsicke, B.~T.\ 2016a, MNRAS, 461, 1413 

\bibitem[Veras et al.(2016b)]{veretal2016b} Veras, D., Mustill, A.~J., G{\"a}nsicke, B.~T., et al.\ 2016b, MNRAS, 458, 3942 

\bibitem[Veras et al.(2017a)]{veretal2017a} Veras, D., Carter, P.~J., Leinhardt, Z.~M., \& G{\"a}nsicke, B.~T.\ 2017a, MNRAS, 465, 1008 

\bibitem[Veras et al.(2017b)]{veretal2017b} Veras, D., Georgakarakos, N., Dobbs-Dixon, I., \& G{\"a}nsicke, B.~T.\ 2017b, MNRAS, 465, 2053 

\bibitem[\protect\citeauthoryear{Veras et al.}{2018}]{veretal2018} Veras D., Georgakarakos N., G{\"a}nsicke B.~T., Dobbs-Dixon I., 2018, MNRAS, 481, 2180

\bibitem[Veras et al.(2019)]{veretal2019} Veras, D., Higuchi, A., \& Ida, S.\ 2019, MNRAS, 485, 708

\bibitem[Veras \& Scheeres(2020)]{versch2019} Veras, D., Scheeres, D., \ 2020, MNRAS, 492, 2437

\bibitem[Veras et al.(2020a)]{veretal2020a} Veras, D., McDonald, C.~H., \& Makarov, V.~V.\ 2020a, MNRAS, 492, 5291

\bibitem[Veras et al.(2020b)]{veretal2020b} Veras, D., Tremblay, P.-E., Hermes, J.~J., et al.\ 2020b, MNRAS In Press, arXiv:2001.08757

\bibitem[Villaver et al.(2014)]{viletal2014} Villaver, E., Livio, M., Mustill, A.~J., \& Siess, L.\ 2014, ApJ, 794, 3

\bibitem[Vokrouhlick{\'y} et al.(2015)]{voketal2015} Vokrouhlick{\'y}, D., Bottke, W.~F., Chesley, S.~R., et al.\ 2015, Asteroids IV, 509

\bibitem[Volk \& Malhotra(2019)]{volmal2019} Volk, K., \& Malhotra, R.\ 2019, AJ, 158, 64

\bibitem[Wang et al.(2015)]{wang2015b} Wang L., Spurzem R., Aarseth S., Nitadori K., Berczik P., Kouwenhoven M. B. N., Naab T., 2015b, MNRAS, 450, 4070

\bibitem[Wang et al.(2016)]{wang2016} Wang L., Spurzem R., Aarseth S.,Giersz M., Askar A., Berczik P., Naab T., Schadow R., Kouwenhoven M. B. N., 2016, MNRAS, 458, 1450

\bibitem[Wang et al.(2020)]{wanetal2020} Wang, Y.-H., Perna, R., \& Leigh, N.~W.~C.\ 2020, Submitted to MNRAS, arXiv:2002.05727

\bibitem[Warner et al.(2009)]{warnetal2009} Warner, B.~D., Harris, A.~W, Pravec, P. \ 2009, Icarus, 202, 134

\bibitem[Weissman \& Levison(1997)]{weilev1997} Weissman, P.~R., \& Levison, H.~F.\ 1997, ApJL, 488, L133

\bibitem[Welsh \& Montgomery(2015)]{welshmontgomery2015} Welsh, B.~Y., \& Montgomery, S.~L.\ 2015, Advances in Astronomy, 2015, 980323

\bibitem[Williams \& Cieza(2011)]{williamscieza2011} Williams, J.~P., \& Cieza, L.~A.\ 2011, ARA\&A, 49, 67

\bibitem[Williams(2017)]{williams2017} Williams, G.\ 2017, MPEC, 2017-U181: : Comet C/2017 U1 (PanStarrs).
IAU Minor Planet Center https://www.minorplanetcenter.net/mpec/K17/K17UI1.html  

\bibitem[Winter et al.(2018)]{winetal2018} Winter, A.~J., Clarke, C.~J., Rosotti, G., et al.\ 2018, MNRAS, 478, 2700

\bibitem[Wisdom \& Holman (1991)]{wisdom1991} Wisdom, J.; Holman, M., 1991, AJ,102, 1528

\bibitem[Wolff et al.(2012)]{woletal2012} Wolff, S., Dawson, R.~I., \& Murray-Clay, R.~A.\ 2012, ApJ, 746, 171

\bibitem[Wyatt et al.(2007)]{wyaetal2007} Wyatt, M.~C., Smith, R., Su, K.~Y.~L., et al.\ 2007, ApJ, 663, 365

\bibitem[Wyatt et al.(2014)]{wyaetal2014} Wyatt, M.~C., Farihi, J., Pringle, J.~E., \& Bonsor, A.\ 2014, MNRAS, 439, 3371 

\bibitem[Wyatt et al.(2017)]{wyaetal2017} Wyatt, M.~C., Bonsor, A., Jackson, A.~P., et al.\ 2017, MNRAS, 464, 3385

\bibitem[Xu et al.(2017)]{xuetal2017} Xu, S., Zuckerman, B., Dufour, P., et al.\ 2017, ApJ, 836, L7

\bibitem[Xu et al.(2019)]{xuetal2019} Xu, S., Dufour, P., Klein, B., et al.\ 2019, AJ, 158, 242

\bibitem[Zakamska \& Tremaine(2004)]{zaktre2004} Zakamska, N.~L., \& Tremaine, S.\ 2004, AJ, 128, 869 

\bibitem[Zeebe(2015)]{zeebe2015} Zeebe, R.~E.\ 2015, ApJ, 798, 8

\bibitem[Zheng et al.(2015)]{zheetal2015} Zheng, X., Kouwenhoven, M.~B.~N., \& Wang, L.\ 2015, MNRAS, 453, 2759

\bibitem[Zotos \& Veras(2020)]{zotver2020} Zotos, E.E., Veras, D., Submitted to A\&A

\bibitem[Zuckerman \& Becklin(1987)]{zucbec1987} Zuckerman, B., \& Becklin, E.~E.\ 1987, Nature, 330, 138

\bibitem[Zuckerman et al.(2003)]{zucetal2003} Zuckerman, B., Koester, D., Reid, I.~N., H\"{u}nsch, M.\ 2003, ApJ, 596, 477 

\bibitem[Zuckerman et al.(2007)]{zucetal2007} Zuckerman, B., Koester, D., Melis, C., Hansen, B.~M., \& Jura, M.\ 2007, ApJ, 671, 872

\bibitem[Zuckerman et al.(2010)]{zucetal2010} Zuckerman, B., Melis, C., Klein, B., Koester, D., \& Jura, M.\ 2010, ApJ, 722, 725 

\end{thebibliography}
\end{document}